\documentclass[aps,prb,twocolumn,floatfix,longbibliography,nofootinbib,superscriptaddress]{revtex4-2}
\usepackage[utf8]{inputenc}
\usepackage{natbib}
\usepackage{graphicx}
\usepackage{xcolor}
\usepackage{tikz}
\usepackage{braket}
\usepackage[tbtags]{amsmath}
\usepackage[colorlinks, linkcolor=blue]{hyperref}
\hypersetup{colorlinks,allcolors=blue}
\usepackage{amssymb}
\usepackage{amsmath}
\usepackage{gensymb}
\usepackage{float}
\usepackage{comment}
\usepackage{amsmath}
\usepackage{tabularx,graphicx}
\usepackage{epstopdf}
\usepackage{latexsym}
\usepackage{color, colortbl}
\usepackage{psfrag}
\usepackage{bbm}
\usepackage{bm}
\usepackage{lmodern}
\usepackage{fix-cm}
\usepackage{xfrac}
\usepackage{color}    
\usepackage{lipsum}
\usepackage{textcomp}
\usepackage{amsmath}
\usepackage{mathtools}
\usepackage{subfigure}
\usepackage{titlesec}
\usepackage{dsfont}
\usepackage{feynmp}
\usepackage{slashed}
\usepackage{multirow}
\usepackage[normalem]{ulem}
\usepackage{svg}
\renewcommand{\vec}[1]{\boldsymbol{#1}}

\newcommand{\bcen}{\begin{center}}
\newcommand{\ecen}{\end{center}}
\newcommand{\btab}{\begin{tabular}}
\newcommand{\etab}{\end{tabular}}
\newcommand{\bdes}{\begin{description}}
\newcommand{\edes}{\end{description}}

\newcommand{\beq}{\begin{equation}}
\newcommand{\eeq}{\end{equation}}
\newcommand{\bea}{\begin{eqnarray}}
\newcommand{\eea}{\end{eqnarray}}

\newcommand{\bary}{\begin{array}}
\newcommand{\eary}{\end{array}}
\newcommand{\benum}{\begin{enumerate}}
\newcommand{\eenum}{\end{enumerate}}
\newcommand{\bitem}{\begin{itemize}}
\newcommand{\eitem}{\end{itemize}}

\renewcommand{\vec}[1]{\boldsymbol{#1}}
\newcommand{\btau}{\mbox{\boldmath $ \tau $}}

\newcommand{\bB} { \mbox{\boldmath $B$}}

\DeclareMathOperator*{\Motimes}{\text{\raisebox{0.25ex}{\scalebox{0.8}{$\bigotimes$}}}}
\newcommand{\Fig}[1]{Fig.~\ref{#1}}

\makeatletter

\newcommand{\Rmnum}[1]{\expandafter\@slowromancap\romannumeral #1@}
\makeatother

\begin{document}

\title{Tunable anyonic permeability across ${\mathbb{Z}_2}$ spin liquid junctions}

\author{Sayak Bhattacharjee}
\altaffiliation{equally contributed}
\email{sayakbhattacharjee@stanford.edu}
\affiliation{Department of Physics, Stanford University, Stanford, California 94305, USA}

\author{Soumya Sur\footnote{equally contributed}}
\altaffiliation{equally contributed}
\email{ssoumya@iitk.ac.in}
\affiliation{Department of Physics, Indian Institute of Technology Kanpur, Kalyanpur, UP 208016, India}

\author{Adhip Agarwala}
\email{adhip@iitk.ac.in}
\affiliation{Department of Physics, Indian Institute of Technology Kanpur, Kalyanpur, UP 208016, India}

\begin{abstract}
We introduce two classes of junctions in a toric code, a prototypical model of a $\mathbb{Z}_2$ quantum spin liquid, and study the nature of anyonic transport across them mediated by Zeeman fields. In the first class of junctions, termed potential barrier junctions, the charges sense effective static potentials and a change in the band mass. In a particular realization, while the junction is completely transparent to the electric charge, magnetic charge transmission is allowed only after a critical field strength. In the second class of junctions, we stitch two toric codes with operators which do not commute at the junction. We show that the anyonic transmission gets tuned by effective pseudospin fluctuations at the junction. Using exact analytical mappings and numerical simulations, we compute charge-specific transmission probabilities. Our work, apart from uncovering the rich physical mechanisms at play in such junctions, can motivate experimental work to engineer defect structures in topologically ordered systems for tunable transport of anyonic particles.

\end{abstract}

\maketitle

\section{Introduction}

Controlled anyonic scattering has been at the forefront of quantum technology research in a wide range of systems---especially in the context of the fractional quantum Hall effect \cite{comforti2002bunching, han2016topological, nakamura2020direct, nature_anyonint, hashisaka2021andreev, doi:10.1126/science.aaz5601, biswas2022shot, kundu2023anyonic, ghosh2024coherentbunching, veillon2024observation} and quantum spin liquids \cite{PhysRevLett.127.167204, PhysRevLett.126.177204, PhysRevX.12.011034, PhysRevB.106.024411, adhip_anyonscattering, PhysRevB.107.104406, PhysRevB.110.214426, PhysRevLett.132.206501, zhou2025probinganyon} in solid-state devices as well as cold-atom simulators \cite{PhysRevLett.115.053002, PhysRevB.92.075116, PhysRevLett.121.030402, gorg2019realization, PhysRevX.10.021031, leonard2023realization, kwan2024realization, shen2025realizationfermioniclaughlinstate}. While the study of disorder \cite{PhysRevLett.104.237203, PhysRevB.92.014403, PhysRevLett.117.037202, PhysRevB.102.054437, PhysRevX.11.011034, PhysRevLett.129.037204, lee2023kondo} and defects \cite{kitaev_toriccodedefeft, bombin_twistdefect, Barkeshli-defect-classification, PhysRevB.90.115118, wen_gappeddomainwalls, PhysRevB.90.134404, peps_domainwalls} in $\mathbb{Z}_2$ spin liquids has been extensively done, the study of junctions of topologically ordered structures and their role in anyonic transport has been little explored \cite{doi:10.1126/science.1253251, PhysRevLett.125.267206, PhysRevB.104.235118, PhysRevResearch.5.013169}. Although junctions have played a defining role in tuning electronic transport, the development of governing principles and/or concrete models for anyonic equivalents is still nascent. 
In this work, we fill this gap by introducing two classes of junctions between otherwise translationally uniform two-dimensional $\mathbb{Z}_2$ toric codes (TC) \cite{kitaev_toriccode}, the paradigmatic model for a $\mathbb{Z}_2$ spin liquid \cite{Savary_2017}.  The ground state(s) of the TC exhibit topological order \cite{RevModPhys.89.041004} whose low-energy excitations are Abelian anyons \cite{leinaas1977theory, wilczek1990fractional, PhysRevLett.67.937, RevModPhys.80.1083}---bosonic electric ($e$) and magnetic charges ($m$) and their fermionic bound state, the dyon ($\psi$). Our focus on this work will be on tuning the transmission probabilities of $e$ and $m$ charges by distinct physical mechanisms dependent on the nature of the two classes of junctions we discuss: (1) the potential barrier junction and (2) the phase junction.

\begin{figure}
\centering
\includegraphics[width=0.9\columnwidth]{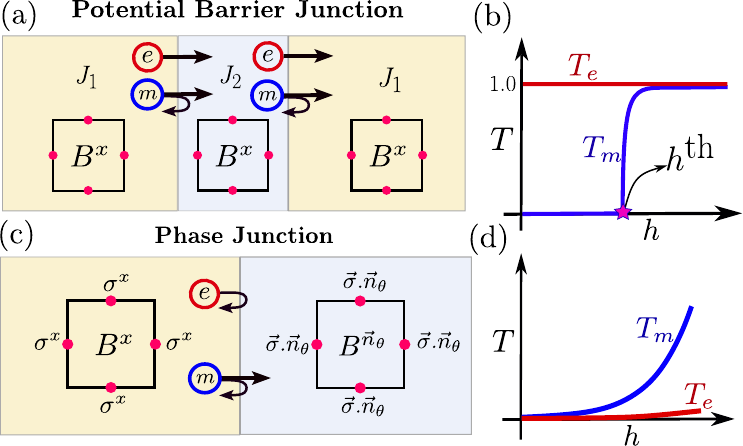}
\centering
\caption{\textbf{Two classes of toric code junctions:} Schematic figures of {\bf(a)} a {\it potential barrier} $J_1$-$J_2$ junction (Eq.~\eqref{eq:J1-J2code}) and {\bf(c)} the {\it phase junction} (Eq.~\eqref{xy_toric code}), with $e$ and $m$ anyon transport mediated by external Zeeman fields ($h$). Junction {\bf(a)} is transparent to $e$ anyons ($T_e=1$) and can switch from being opaque ($T_m = 0$) to transparent for $m$ anyons at a threshold field strength $h^{\rm{th}}$ (purple star in \textbf{(b)}). Junction {\bf(c)} is opaque to both $e$ and $m$ due to strong quantum fluctuations at the junction from effective pseudospins (see main text).}
\label{fig_schematic}
\end{figure}

In the {\it potential barrier junction} we uniformly modify the TC couplings in a region sandwiched between two other regions. We consider the TC defined on a square lattice \cite{kitaev_toriccode} with two sets of stabilizers---star ($A^z_v$) and plaquette ($B^x_p$), associated with a vertex $v$ and plaquette $p$, respectively---each constructed using spin-$1/2$ ($\bm{\sigma}$) operators of strength $J$ defined on the links, and with a Hamiltonian,
\begin{equation} \label{eq:TC}
H_{\rm{TC}}=-J\sum_{v}A^{z}_{v} - J \sum_{p} B^{x}_{p}
\end{equation}
Excitations in $A^z_v$ are referred to as $e$ charges, while those in $B^x_p$ are called $m$ charges (note that we use a unitarily equivalent definition compared to \cite{kitaev_toriccode}). We construct a $J_{1}$-$J_{2}$ junction (J1J2) (see Fig.~\ref{fig_schematic}(a)), where we modify the coupling of $B^x_p$ in a finite intermediate region to $J_2$ while the rest of the system has coupling $J_1$. This renders the junction semi-permeable to tunneling by an $m$ anyon, while it is transparent to an $e$ anyon. A Zeeman field of strength $h$ can selectively induce $e$ or $m$ particles to tunnel across the junction. While the transmission probability for the $e$ charge $T_e=1$, the magnetic charge $m$ transmits only beyond a threshold value of $h=h^\textrm{th}$ (see Fig.~\ref{fig_schematic}(b)). This should be contrasted with the {\it phase junction}, where we stich two otherwise homogeneous TCs, but where in the second region the $B^x_p$ operators are rotated to an arbitrary spin direction $\vec{n}_\theta = (\cos{\theta}, \sin{\theta},0)$, $\theta\in (0,2\pi)$ (see Fig.~\ref{fig_schematic}(c)) such that $B_p^{\vec{n}_\theta}= \Motimes_{l\in p}\vec{\sigma}_{l}\cdot\vec{n}_{\theta}$ ($l$ denotes a link of the lattice). This generically results in a non-commuting junction between the two regions. In this case, even in the absence of any inhomogeneous coupling strengths, the junctions show remarkable opacity to both $e$ and $m$, which, however, are smoothly tunable by field strength ($h$) (see Fig.~\ref{fig_schematic}(d)) and angle ($\theta$). As we will show, both junctions have distinct physical mechanisms leading to their charge-specific permeability.

In Section~\ref{pb-junction}, we discuss the transmission properties of the potential barrier junction. We introduce the $J_1$-$J_2$ model of the TC (Sec.~\ref{j1j2-model}) and derive an effective spin chain model to describe the field-induced anyon dynamics (Sec.~\ref{sec-j1j2-map-to-ising}). The results for anyon transmission are discussed in Sec.~\ref{tr-calc}. Next, we introduce the phase junction model of the TC in Sec.~\ref{pjunction} and discuss its properties for general phase $\theta$. In Sec.~\ref{xyjc-model}, we consider a specific limit ($\theta=\pi/2$) of the phase junction, which we term the
XY junction (XYJC) model, and discuss the ground state and quasiparticle excitations at zero Zeeman fields in Sec.~\ref{xyjc-gs-exc}. In Sec.~\ref{m-charge-scattering}, and Sec.~\ref{e-charge-scat}, we discuss the scattering and transmission properties of $m$ and $e$ anyons across the XYJC model. Next, in Sec.~\ref{res-gen-theta}, we discuss the dependence of transmission properties on $\theta$, for $\theta$ values close to the anti-commuting XYJC limit. Finally, Sec.~\ref{sec-coclude} concludes the paper and outlines some future directions.


\section{Potential barrier junction} \label{pb-junction}

\subsection{$J_1$-$J_2$ model} \label{j1j2-model}

The Hamiltonian is given by 
\begin{equation} \label{eq:J1-J2code}
H_{\text{J1J2}}=-J\sum_{v}A^{z}_{v} - \sum_{p}J(\vec{p})B^{x}_{p}
\end{equation}
where $B_p^\alpha =  \Motimes_{l\in p}\sigma^\alpha_{l}$ and $A_v^\alpha = \Motimes_{l\in v}\sigma^\alpha_{l}$ are the plaquette $p$ and vertex $v$ operators respectively (and $l$ denotes a link on the lattice), $\sigma^\alpha$ ($\alpha=x,y,z$) are the Pauli operators. Here, $\vec{p} = \{p_x, p_y\}$ denotes the position of a plaquette and $J,J(\vec{p})>0$. 
While everywhere else $J(\vec{p})=J_1$, in an intermediate region of width $W$ (indexed {$p_x \in \{1, W\}, \forall\: p_{y}$) $J(\vec{p}){=} J_2$ (see Fig.~\ref{fig:J1_J2}(a)). The topologically ordered ground state(s) and the excitations of the standard TC ($J_2=J_1$) remain stable when $J_2 \neq J_1$. Note that even though all operators mutually commute, crucially, the energy gap for $m$ anyons differs between the $J_1$ and $J_2$ regions.

\begin{figure}
\centering
\includegraphics[width=1\columnwidth]{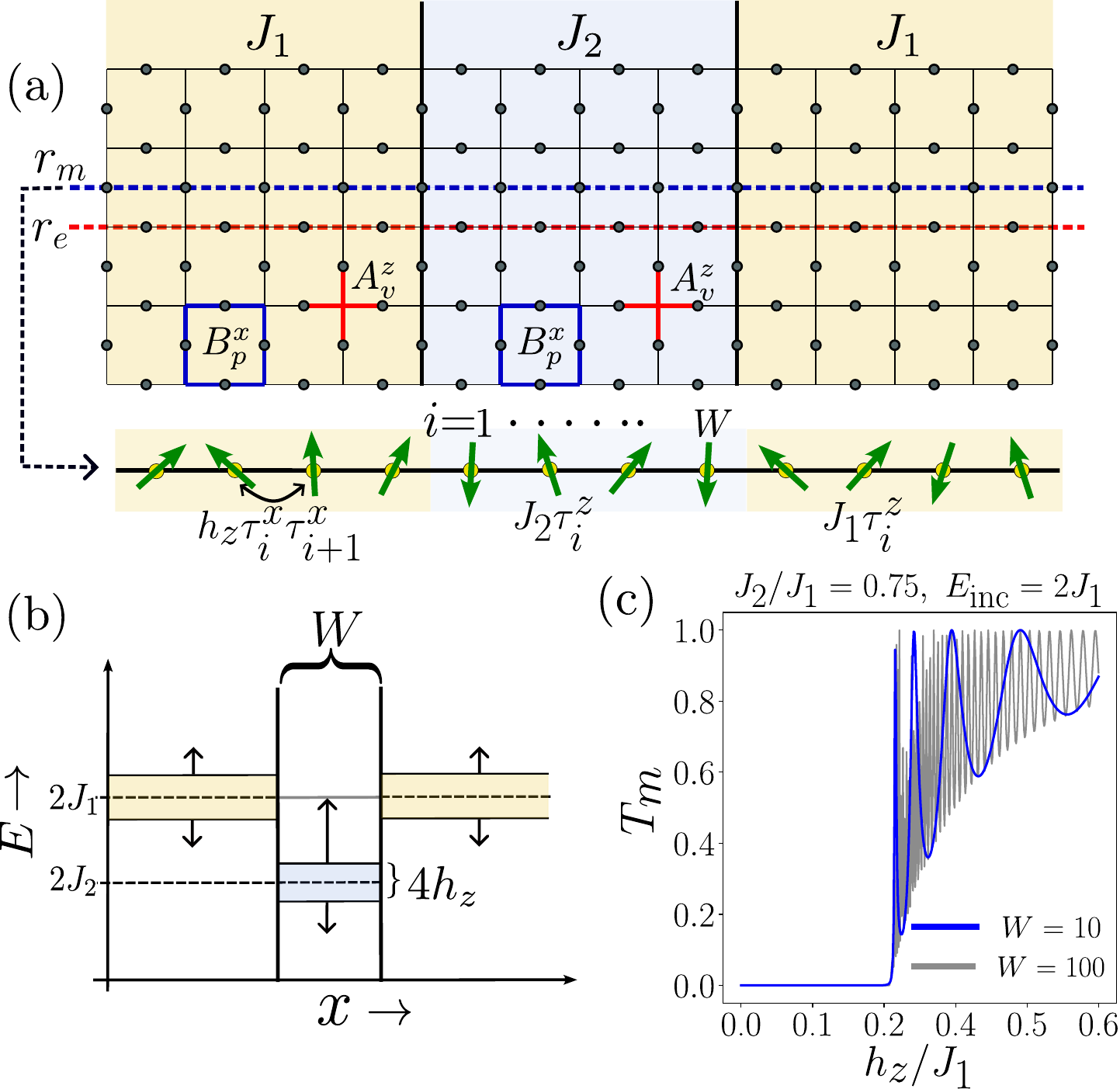}
\centering
\caption{\textbf{The $J_1$-$J_2$ TC junction}: {\bf(a)} Schematic of lattice model (Eq.~\eqref{eq:J1-J2code}) with star ($A_{v}^{z}$) and plaquette ($B_{p}^{x}$) operators. Dynamics of $e,m$  charges induced by Zeeman field along paths $r_{e}$ and $r_{m}$ can be mapped to a transverse field Ising chain (Eq.~\eqref{dual-spin-H-J1-J2}). {\bf(b)} $h_{z}$ broadens the $m$ bands leading to a band-mismatch when $J_2\neq J_1$.  {\bf(c)} Transmission probability ($T_{m}$) of $m$ anyons as a function of field ($h_{z}/J_{1}$) shows a jump at a threshold value $h_z= h^{\rm{th}}_z$ which is independent of junction size $W$ (data for $E_{\rm{inc}}=2J_{1}$, $J_{2}/J_{1}=0.75$, $J_1=1$).}
\label{fig:J1_J2}
\end{figure}

 Anyon selective dynamics can be induced by applying an external Zeeman field of strength $h_\alpha$ on specific one-dimensional paths such that $H=H_{\rm{J1J2}}+H_f^\alpha$ where $H_f^\alpha = -h_\alpha\sum_l\sigma_l^\alpha$. While a dispersive $e$ charge can be moved by applying an $x$-directional field along row $r_e$ ($\alpha=x$) passing through horizontal links, an $m$ charge can be equivalently moved by action of $h_z$ ($\alpha=z$) on a path $r_m$ cutting the vertical links as shown in Fig.~\ref{fig:J1_J2}(a)).  We focus on the limit $h_\alpha/J_1 \ll 1$, ensuring a low-density regime in which anyons are sparsely distributed and associated fluctuations are below the flux gap, allowing us to study their transport at the single-particle level.

\subsection{Mapping to the transverse field Ising model} \label{sec-j1j2-map-to-ising}


The strip of the $J_1$-$J_2$ TC in which the magnetic fields act can be each mapped to a transverse field Ising model (TFIM)  \cite{PhysRevLett.98.070602, PhysRevB.83.075124, PhysRevB.100.125159}. 
The plaquette and star operators are Ising-like variables (since $(A_v^z)^2=(B^x_p)^2=1$), located at the vertices and plaquette centers of the square lattice, respectively. Therefore, $A_{v}^{z}\mapsto \tau^{z}_{v}$, and $B_{p}^{x}\mapsto \tau^{z}_{p}$, where $\tau^{z}_{p/v}$ are Pauli operators with eigenvalues $\pm 1$. Since $\lbrace A^z_{v}, \sigma^x_l\rbrace=\lbrace B^x_{p}, \sigma^z_l\rbrace=0$, whenever $l\in v / p$ and commute otherwise, the field along the $x$ ($z$)-direction only causes fluctuations in the $e$ ($m$) sector via either anyon hopping or pair creation; hence, its effective action in terms of `dual' $\tau$ spins is $\sim h\tau^{x}_{i}\tau^{x}_{i+1}$, where $i=v/p$ (see Appendix \ref{Mapping-J1-J2-model-to-Ising} for the full derivation). In terms of the $\btau$ spins, 
\begin{equation}
H^{(q)} = -\sum_{i\in r_{q}}J_{q}(i)\tau^{z}_{i} - h_{\alpha}\sum_{i \in r_{q}}\tau^{x}_{i}\tau^{x}_{i+1} \label{dual-spin-H-J1-J2}
\end{equation}
where $H^{(q)}$ is the effective Hamiltonian describing the dynamics of anyon $q=e,m$. Since $h_{x}$ ($h_{z}$) along $r_{e}$ ($r_{m}$) leads to dynamics only in the $e$ ($m$) sector, we use the common label $\alpha$ to differentiate them. $J_{m}(i\in r_{m})=J_{2}$ for $1\leq i\leq W$ and for the rest of the chain $J_{m}=J_{1}$. $J_{e}(i)=J$ throughout the chain $r_{e}$ (see lower panel in Fig.~\ref{fig:J1_J2}(a)).

Applying the standard Jordan-Wigner (JW) fermionization followed by a Bogoliubov transformation in the quasi-momentum ($k$) basis yields the energy dispersion $(E_k)$ of the uniform Ising chain \cite{sachdev_QPT}. For $h_{\alpha}/J_{1}\ll 1$ and near the band minimum $k=0$, $E^{q}_{k}\approx \Delta_{q} + k^{2}/2M_{q}$. Here, $\Delta_{q}= 2|J-h_{\alpha}|$ is the energy gap and $M_{q} = |(1-h_{\alpha}/J)|/2h_{\alpha}$ is the band mass of an anyon. At low energies, anyon propagation (with wavefunction $\psi_q$) is governed by the Schrodinger equation: $-\partial_{x}^{2}\psi_q/2M_q(x) + (V_q(x)-E)\psi_{q}=0 $ where $E$ is the energy. The $e$ anyon Hamiltonian exhibits translational symmetry: $M_{e}(x)$ is constant, and the potential $V_{e}(x)=0$ throughout the system. Thus, the $e$ anyons freely disperse above the energy gap and $T_{e}(E)=1$ for any $h_{x}$ and incident energy $E_{\rm{inc}}$. For the $m$ anyon, $V_{m}(x)=2(J_{2}-J_{1})$ and $M_{m}(x)$ changes when crossing the $J_{1}$-$J_{2}$ boundary; the $m$ anyon is lighter if $J_{2}<J_{1}$ and heavier otherwise. Therefore, the $m$ anyon experiences a junction realizing a {\it potential barrier/well} with the band mass concomitantly changing across the junction. Moreover, $h$ decides the effective bandwidth of the anyonic bands centered around the bare flux-gap, which is decided by the $J_1$ or $J_2$ scale. Thus, tuning $h$ and $J_2 - J_1$ one can mimic the band-pass filter physics of semiconducting systems (see Fig.~\ref{fig:J1_J2}(b)) but now for anyonic transport.  

\subsection{Transmission calculation} \label{tr-calc}

We numerically compute the $m$-anyon transmission probability $T_{m}(E,W)$ for arbitrary $h_{z}$ and $E_{\rm{inc}}$ using a discretized wavefunction matching (WFM) approach \cite{kelly_transmission_calculation} to link eigenstates of the $J_{1}$ (leads) and $J_{2}$ (scatterer) regions in the JW fermionized Eq.~\eqref{dual-spin-H-J1-J2} (Also see Appendix \ref{wavefunction-matching-transmission} for a related discussion in the context of XYJC). The main results are summarized in Fig.~\ref{fig:J1_J2}(c). For a fixed $J_2/J_1=0.75$ (potential well) and for $E_{\rm{inc}}=2J_{1}$ (the zero-field gap of $m$ anyon), we plot $T$ for two different widths ($W=10$ and $W=100$) of the $J_{2}$ region. Beyond a threshold $h^{\rm{th}}_{z}\approx 0.25J_1$, transmission occurs only through {\it resonant tunneling}. With increasing $W$, the number of peaks increases (Fig.~\ref{fig:J1_J2}(c)) as more discrete energy levels (of the $J_{2}$ region) match $E_{\rm{inc}}$. For $E_{\rm{inc}}=2J_1$, threshold value $h^{\rm{th}}_{z} = J_1-J_2$ (for $J_2<J_1$), which is when the upper-band edge of the intermediate region hits the band-bottom of the $J_1$ regions (also see Fig.~\ref{fig:J1_J2}(b) consistent with this argument). Thus $E_{\rm{inc}}$ and $h_z$ serve as tuning parameters to modulate anyonic permeability across the junction.

\begin{figure}
\centering
\includegraphics[width=1.0\columnwidth]{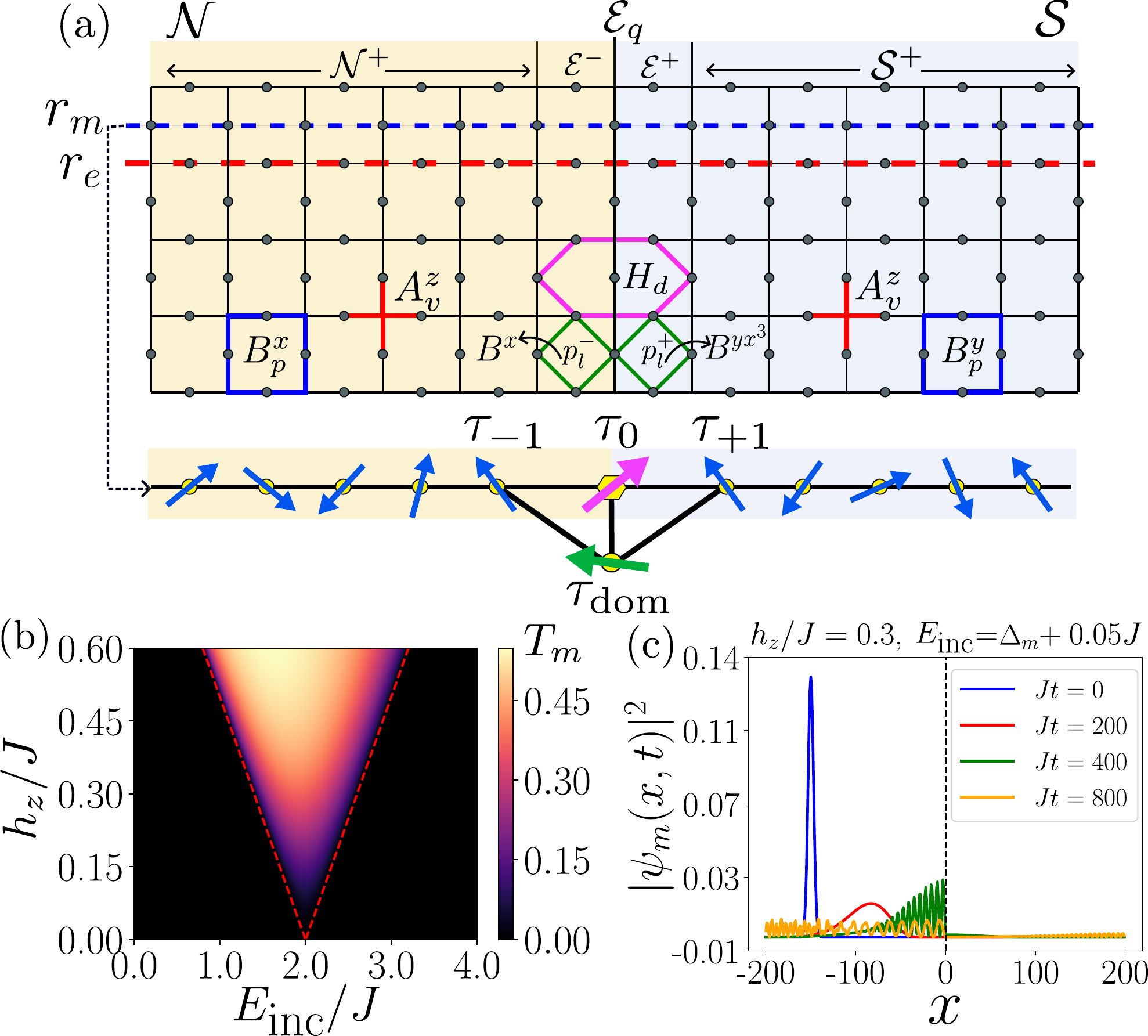}
\centering
\caption{\textbf{The XYJC junction:} {\bf (a)} The 2D lattice Hamiltonian with its constituent star, plaquette, domino and dimer operators (see Eqs.~\eqref{xy_toric code},\eqref{eq_rotatedquantumjunction}). Dynamics induced by a Zeeman field along path $r_{m}$ can be mapped to an Ising chain with a three-spin coupling mediated by an effective pseudospin $\tau_{\text{dom}}$. {\bf(b)} Transmission probability ($T_m$) for a magnetic charge as a function of field $h_z$ and incident energy $E_{\rm{inc}}$. The anyon band edges are shown by red dashed lines. {\bf (c)} Time evolution snapshots of an $m$ anyon wave packet ($\psi_{m}(x,t)$) thrown from the $\mathcal{N}^{+}$ region towards the junction at different times, showing effective obstruction due to pseudospin ($h_z=0.3J,E_{\rm{inc}}= \Delta_{m}+0.05J$, $J=1$, $\Delta_{m}=2(J-h_z)$). }
\label{fig:xy-toric-code}
\end{figure}

\section{Phase junction} \label{pjunction}

We next consider another class of a TC junction, where two otherwise uniform TCs are stitched to form a non-commuting junction. In the first region (labeled $\mathcal{N}$ for north) we take the usual TC (see Eq.~\eqref{eq:TC}), while in the second region (labeled $\mathcal{S}$ for south) the $B^{x}_{p}$ operators are spin-rotated to $B^{\vec{n}_{\theta}}$ \cite{junction_notation}. The Hamiltonian is given by,
\begin{align}
H_{\theta\text{JC}}=-J\sum_{p\in \mathcal{N}}B_{p}^{x}-J\sum_{p\in \mathcal{S}}B_{p}^{\vec{n}_{\theta}}-J\sum_{v}A_{v}^{z}
\end{align}
Note that the set of links ($l$) on the (vertical) junction line---which we call the equator $\mathcal{E}_q$ ---are shared between the two types of $B_p$ operators defined on the two plaquettes, $ p^{\pm}_l$ (see Fig.~\ref{fig:xy-toric-code}(a)), which do not commute: $[B^x_{p_l^-}, B^{\vec{n}_{\theta}}_{p_{l}^+}] \neq 0$ for $\theta \neq 0,\ l\in \mathcal{E}_{q}$. When $\theta=\pi/2$,  $\{B_{p_l^-}^x, B_{p_l^+}^y\}=0$ \cite{commfoot}, and we refer to this as the {\it XY junction} (XYJC), which will be studied in detail in the next subsection. Before that, we discuss a few properties of $H_{\theta \text{JC}}$ for general $\theta$. 

It is useful to segregate the microscopic spin operators $\sigma^\alpha_l$ on the links $l$ into three regions $\{\mathcal{N},\mathcal{S}, \mathcal{E}_q\}$ depending upon which region they occur. Similarly, the plaquette operators can be conveniently divided into sectors $\{\mathcal{N}^+, \mathcal{S}^+\}$ when all their links exclusively belong to either $\mathcal{N}$ or $\mathcal{S}$ respectively; and $\{\mathcal{E}^-, \mathcal{E}^+\}$ when they share a link with the $\mathcal{E}_q$ line (see Fig.~\ref{fig:xy-toric-code}(a)). We rotate all spin operators on the right of $\mathcal{E}_q$ line by, 
\begin{align}
U(\theta) = \prod_{l\in \mathcal{S}} e^{-i \sigma^{z}_{l}\theta/2}
\end{align}
Therefore, the Hamiltonian transforms to $U^{\dag}H_{\theta\text{JC}} U =\tilde{H}_{\theta\text{JC}}$,
\begin{align}
\tilde{H}_{\theta\text{JC}} = -J\sum_{p\in \mathcal{N},\ \mathcal{S}^{+}}B_p^x-J\sum_{p\in \mathcal{E}^+}B_p^{x^3 n_{\theta}}-\sum_vA_v^z
\end{align}
where $B_p^{x^3 n_{\theta}}= U{B_p^y}U^\dag =(\vec{\sigma}_{l_{\mathcal{E}}}\cdot\vec{n}_{\theta})\Motimes_{l\in \mathcal{S}, p} \sigma^x_{l}$ and $l_\mathcal{E}$ is a link on the equator. 

We now introduce the \textit{domino} operator, $D_{d}(\theta)=(B_{p_{l}^{-}}^{x}+B^{x^{3}n_{\theta}}_{p^{+}_{l}})/\sqrt{2}$ to write $\tilde{H}_{\theta\text{JC}}$ as the following,
\begin{align}
\tilde{H}_{\theta\text{JC}}= -J\sum_{p\in \mathcal{N}^+, \mathcal{S}^+}B_p^x-\sqrt{2}J\sum_{d\in \mathcal{E}_q}D_d(\theta)-\sum_vA_v^z     \label{theta-junction-zero-h}
\end{align}

Here, $d$ denotes the rectangular two-plaquette ($p_{l}^- \cup p_l^+$) centered on a link $l_{\mathcal{E}}$ lying on the equator ($\mathcal{E}_q$). The above Hamiltonian also has a local symmetry; the hexagonal-shaped \textit{dimer}, $H_d=B_{p_l^-}^x\otimes B_{p_l^+}^{x}$ commutes with $\tilde{H}_{\theta\text{JC}}$ (see \Fig{fig:xy-toric-code}(a)). The domino operators satisfy the following algebraic relations: (1) $D_{d}^{2}(\theta)= (1+H_{d}\cos{\theta}) $, (2) $[D_{d_{1}}(\theta_{1}), D_{d_{2}}(\theta_{2})]=0$ for any two domino plaquettes $d_{1}, d_{2} \in \mathcal{E}$, and (3) $[D_{d}(\theta), B_{p}^{x}]=[D_{d}(\theta), A_{v}^{z}]=0$. Thus, all the {\it stabilizer} operators commute, and the exact ground state is obtained by choosing the maximum eigenvalues of these stabilizer operators (Eq.~\eqref{theta-junction-zero-h}). When $0\leq \theta < \pi/2$, to maximize $D_d(\theta)$ eigenvalues, we must have $H_{d}=+1$ for all the rectangular plaquettes ($d\in \mathcal{E}_q$). As a result, the ground state corresponds to a single uniform configuration of $H_d=+1$. A single dimer-flip excitation, obtained by flipping one $H_{d}$ eigenvalue, has an energy gap,
\begin{align}
\Delta_{H}(\theta) &= E_{1H}(\theta) -E_{0}(\theta) \nonumber\\
&= \sqrt{2}J\big( \sqrt{1+\cos{\theta}} - \sqrt{1-\cos{\theta}} \big) \label{vortex-gap}
\end{align}
At the zero-field level, these excitations are immobile, like the usual $e$ and $m$ charges. The dimer-flip gap ($\Delta_H(\theta)$) plays an important role in the $m$ anyon transmission properties when $\theta\neq \pi/2$. Before discussing this, we discuss the physics of XYJC model ($\theta=\pi/2$).

\subsection{XY junction code model} \label{xyjc-model}

The Hamiltonian for the anti-commuting XY junction is described as follows,
\begin{align}\label{xy_toric code}
H_{\rm{XYJC}} = -J\sum_{p\in \mathcal{N}}B_p^x-J\sum_{p\in \mathcal{S}}B_p^y-J\sum_vA_v^z,
\end{align}
As before, we also rotate $H_{\rm{XYJC}}$ by, 
\beq
U = \prod_{l\in \mathcal{S}}e^{-i\frac{\pi}{4}\sigma^{z}_{l}} \label{unitary1},
\eeq 
which rotates all  $B^y_p$ operators in the south to $B^x_p$, except the ones in the $\mathcal{E}^+$ region. The  transformed Hamiltonian thus becomes
\begin{equation}
\tilde{H}_{\rm{XYJC}} = -J\sum_{p\in \mathcal{N}^+, \mathcal{E}^-, \mathcal{S}^{+}}B_p^x-J\sum_{p\in \mathcal{E}^+}B_p^{yx^3}-\sum_vA_v^z\label{Eq-byx3-xyjc},
\end{equation}
where $B_p^{yx^3}= U{B_p^y}U^\dag =\sigma^y_{l_{\mathcal{E}}}\Motimes_{l\in \mathcal{S}, p} \sigma^x_{l}$. 

Again, we introduce the composite Ising-like domino operator: $D_{d}=(B_{p_l^-}^x+B_{p_l^+}^{yx^3})/\sqrt{2}$ (see \Fig{fig:xy-toric-code}(a)), and write Eq.~\eqref{Eq-byx3-xyjc} as a {\it frustration-free} Hamiltonian,
\begin{equation}\label{eq_rotatedquantumjunction}
\tilde{H}_{\rm{XYJC}} = -J\sum_{p\in \mathcal{N}^+, \mathcal{S}^+}B_p^x-\sqrt{2}J\sum_{d \in \mathcal{E}_q}D_d-\sum_vA_v^z,
\end{equation}
The non-commutativity is absorbed in $D_{d}$. As a result, (1) $[D_{d_{1}}, D_{d_{2}}]=0$ for any two domino plaquettes $d_{1}, d_{2} \in \mathcal{E}_{q}$, and (2) $[D_{d}, B_{p}^{x}]=[D_{d}, A_{v}^{z}]=0$.
The Hamiltonian (Eq.~\eqref{eq_rotatedquantumjunction}), now rewritten as a set of commuting stabilizers, is tenable to studying the anyon dynamics, discussed next.


\subsection{XY junction code ground state and tower of excitations}\label{xyjc-gs-exc}
The ground state of XYJC is a simultaneous $+1$ eigenstate of $B^x_{p}$ (for $p\in \mathcal{N}^+, \mathcal{S}^+$), $A^z_v$ ($\forall\ v$), and $D_d$ (for $d\in \mathcal{E}_q$). In addition, the {\it dimer} operator $H_d$, which is a symmetry of the XYJC Hamiltonian (Eq.\eqref{eq_rotatedquantumjunction}), results in an additional labeling (quantum number) of the degenerate ground states. For XYJC, a single {\it dimer} excitation, obtained by flipping one $H_{d}$ eigenvalue ($=(-1)^{r^{h}_{d}}$, where $r_{d}^{h}=0/1$), does not cost any energy ($\Delta_H(\theta=\pi/2)=0$).  This enables the ground state to host an arbitrary number of these quasiparticles. This leads to a substantial ground state degeneracy ($\sim 2^{L_{y}}$) for an infinite system, with the degenerate manifold defined by the set $\lbrace r^{h}_{d}\rbrace$, $\forall\ d\in \mathcal{E}_{q}$. These states are given by,

\begin{align}
\big| \textrm{GS},\ \lbrace r^{h}_{d}\rbrace_{d\in \mathcal{E}_{q}}\big\rangle_{\text{XYJC}}=\prod_{d\in \mathcal{E}_{q}}\big(1+(-1)^{r^{h}_{d}}H_{d}\big)\big(1+D_{d}\big)\nonumber\\
\times\prod_{p\in \mathcal{N}^{+}, \mathcal{S}^{+}}(1+B^{x}_{p})\bigotimes_{l} \ket{\uparrow}^{z}_{l}
\end{align}

If the model is defined on a torus (i.e., there are two junctions between the {\it rotated} and {\it unrotated} regions), there is an additional restriction on the allowed configuration of $\lbrace r^{h}_{d}\rbrace$. This is due to the constraint,
\begin{align}
 \prod_{d\in \mathcal{E}_{q}}H_{d}=\prod_{p\in \mathcal{N}^{+}, \mathcal{S}^{+}}B_{p}^{x} \label{relation-betwn-H-and-Bp}   
\end{align}
which implies that the net parity of dimer excitations along the equators (two junctions) is fixed by the total $m$ charge parity in the rest of the system. The ground state(s) should be free of $m$ charges inside $\lbrace \mathcal{N}^{+}, \mathcal{S}^{+}\rbrace$,  which implies $\prod_{d\in \mathcal{E}_{q}}H_{d}\ket{\textrm{GS}}=(+1)\ket{\textrm{GS}}$. Nevertheless, the XYJC ground state has a finite density of $m$ anyons within $\mathcal{E}^{-}\cup\mathcal{E}^{+}$, due to operator non-commutativity (see Appendix \ref{xyjc-appendix-gs-prop})). Hence, the physical ground states (on a torus) are given by,  
\begin{align}
\bigg|\textrm{GS},\ \lbrace r^{h}_{d}\rbrace_{d\in \mathcal{E}_{q}}\ &,\  e^{i\pi\sum_{d}r^{h}_{d}}=1 \bigg\rangle_{\text{XYJC}}^{\text{Torus}} \nonumber\\
&= \big(1+\prod_{d\in \mathcal{E}_{q}}H_{d}\big)   \big| \textrm{GS},\ \lbrace r^{h}_{d}\rbrace_{d\in \mathcal{E}_{q}}\big\rangle_{\text{XYJC}} \label{phys-gs-xytc}
\end{align}
Therefore, apart from the 4-fold topological degeneracy (see Appendix \ref{xyjc-appendix-gs-prop}) characterized by highly non-local \textit{Wilson loops}, the XYJC ground state has an additional $2^{2L_{y}-1}$-fold degeneracy due to local symmetries, where $2L_{y}$ is the total length of the equator ($\mathcal{E}_{q}$) for the lattice with periodic boundary conditions \cite{note-pbc}. Physically speaking, $H_d$ measures $m$ anyon parity within a rectangular plaquette $d$, and the ground state (Eq.~\eqref{phys-gs-xytc}) can have arbitrary local parity fluctuations as long as the global even parity constraint is maintained. 

Apart from the gapless dimer fluctuations, the zero-field XYJC model has conventional static $e$ anyon excitations, which correspond to flipping $A_v^z$ for all the vertices and static $m$ anyon excitations, which correspond to plaquette ($B^x_p$) violations away from the $\mathcal{E}_q$ region. Both of these excitations cost an energy $\sim 2J$. Along the equator, there is a gapped excitation (gap $=2\sqrt{2}J$) generated by flipping the domino operator. As we discuss in the subsequent section, the domino fluctuations play a crucial role in $m$ anyon transport across the junction. 

\subsection{Magnetic $(m)$ charge scattering} \label{m-charge-scattering}

As in the $J_1$-$J_2$ model, to generate dynamics of the $m$ quasiparticles, we apply a Zeeman field $h_z$ along a path $r_m$ which passes through the vertical bonds (see Fig.~\ref{fig:xy-toric-code}(a)). The unitary transformation (Eq.~\eqref{unitary1}) has no effect on the field. While the $z$-field coupling $(h_{z}/J)$ is the same in both the $\mathcal{N}$ and $\mathcal{S}$ regions, operator non-commutativity along the junction leads to a non-trivial tunneling action. To understand the field-induced dynamics, we again go to an Ising model-like description (details in Appendix \ref{field-induced-dynamics-m-anyon-supp}). We replace $B^{x}_{p} \mapsto \tau^{z}_{p}\ (\forall\  p \in \mathcal{N}^{+}, \mathcal{S}^{+})$, $D_{d} \mapsto \tau^{z}_{\text{dom}}$ and $H_d \mapsto \tau^z_0$. Away from $\mathcal{E}^{\pm}$, $\sigma^{z}$ causes either $m$ anyon hopping or pair creation/annihilation, while in the junction region, it also creates fluctuations in the $D_{d}$ and $H_{d}$ sectors, leading to interesting pseudospin ($\tau_{\rm{dom}}$) dynamics, which in turn, influences anyon transmission.

The $\btau$-spin Hamiltonian is given by,
\beq
\tilde{H}^{(m)}=H_{m}+H_{\text{dom}}+H_{I}^{(1)}+H_{I}^{(2)},
\label{mXYJC}
\eeq
where 
\begin{align}
&H_{m}=-J\sum_{i \neq 0} \tau^{z}_{i} - h_{z}\sum_{i\not \in \lbrace -1, 0\rbrace} \tau^{x}_{i}\tau^{x}_{i+1} \\
&H_{\text{dom}}=-\sqrt{2}J\tau^{z}_{\text{dom}}-h_z \tau^{x}_{\text{dom}}, \\
&H_{\textrm{I}}^{(1)} = -\frac{h_{z}}{2}\big[\tau^{x}_{-1}(\tau^{x}_{0}-\tau^{y}_{0})+(\tau^{x}_{0}+\tau^{y}_{0})\tau^{x}_{+1}\big], \\
& H_{\textrm{I}}^{(2)}=-\frac{h_{z}}{2}\big[\tau^{x}_{-1}(\tau^{x}_{0}+\tau^{y}_{0})+(\tau^{x}_{0}-\tau^{y}_{0})\tau^{x}_{+1}\big]\tau^{x}_{\text{dom}}  \end{align}
This describes a TFIM of $\tau_i$ spin-$\frac{1}{2}$s, coupled to an auxiliary pseudospin $\tau_{\rm{dom}}$ at indices $i=0,\pm 1$ through three-body terms (see Fig.~\ref{fig:xy-toric-code}(a), lower panel).

A JW transformation on $\tau_i$, leads to an effective model of a $p$-wave superconductor coupled to an auxiliary $\tau_{\rm{dom}}$ at the center, amenable to a WFM formalism (see Appendix \ref{wavefunction-matching-transmission} for details). This allows us to again numerically calculate the transmission probabilities for the $m$ anyons.  

The results are depicted in Fig.~\ref{fig:xy-toric-code}(b). We find that even at a field of $h_z \sim 0.1 J$, at an incident energy $E_{\rm{inc}} \sim 2J$ (i.e. at the zero-field flux gap) the transmission probability is small ($T_m \sim 0.1$). This can be monotonically increased by increasing $h_z$. This high opacity of the junction is surprising given that, unlike in the J1J2 junction, there is no mismatch at the junction in the coupling strengths. To further verify this, we numerically simulate the time evolution of a Gaussian $m$ anyon wave packet $\psi_m$ with the lattice Hamiltonian $\tilde{H}^{(m)}$ (Eq.~\eqref{mXYJC}) for $E_{\rm{inc}} \sim \Delta_m$ and $h_{z}/J \sim 0.3$. As shown in Fig.~\ref{fig:xy-toric-code}(c), a significant fraction of the wave packet amplitude $|\psi_{m}|^{2}$ is reflected back to the $x<0$ region after hitting the boundary around $Jt \approx 400$.

We next show that this behavior is a direct result of pseudospin ($\vec{\tau}_{\rm{dom}}$) fluctuation. A low-energy ($E\sim \Delta_{m}$) effective theory of $m$-anyon $(\psi)$ dynamics with Hamiltonian (Eq.~\eqref{mXYJC}) (see derivation in Appendix \ref{continuum-model-m-anyon}) is,
\begin{align}
\tilde{H}^{(m)} &\sim -\int dx\ \psi^{\dag}(x)\bigg[\Delta(x) +\frac{\partial_{x}^{2}}{2M_m}\bigg]\psi(x) - \bB_{\rm{eff}} \cdot \vec{\tau}_{\rm{dom}} \nonumber\\ &+  \eta h_{z} \psi^{\dag}(0)\psi(0)\tau^{x}_{\text{dom}},
\end{align}
where $\Delta(x)=\Delta_{m}(1-\delta(x))$, $\bB_{\rm{eff}} =\{h_z,0 ,\sqrt{2}J\}$, $M_m$ is the band-mass ($\propto h_z^{-1} $) and $\eta$ is the lattice-regularization coefficient. At $h_z=0$, as expected, this describes an infinitely heavy particle which is decoupled from a $z$-polarized $\vec{\tau}_{\rm{dom}}$ pseudospin. However, as $h_z$ is applied, it induces perturbative $x$-fluctuations in an otherwise rigid $z$-polarized $\vec{\tau}_{\rm{dom}}$. This leads to a weak transmission of $m$-anyons whose density at $x=0$ is determined by both the onsite potential $\Delta(x)$ and fluctuations in $\vec{\tau}_{\rm{dom}}$. In our wave packet simulation we indeed find that a finite, but small transmission of $\psi_m$ is captured in $\langle \tau_{\rm{dom}}^x\rangle \neq 0$ (see Fig.~\ref{fig:domino-spin-avg} in Appendix \ref{continuum-model-m-anyon}), thus leading to a significant opacity for the $m$ anyons.

\subsection{Electric ($e$) charge scattering} \label{e-charge-scat}

We now discuss the $e$ scattering. As in the $J_1$-$J_2$ model, the Zeeman field ($\propto  h_{x}$) along a path $r_e$ induces $e$ charge dynamics (see \Fig{fig:xy-toric-code}(a)). Under the unitary transformation (Eq.~\eqref{unitary1}) a uniform $x$-field along $r_e$ is mapped to $ h_{x}\sum_{l\in r_{e}}[\delta (l\in \mathcal{N})\sigma^{x}_{l} - \delta (l\in \mathcal{S})\sigma^{y}_{l}]$, whose action is to be studied on the Hamiltonian in Eq.~\eqref{eq_rotatedquantumjunction}. A mapping to TFIM, as before, results in two distinct processes in the north and south regions (see Appendix \ref{field-dynamics-m-anyon-derivation}),

\begin{equation}
H_f^x \mapsto  \begin{cases}
 h_x\sum_i\tau^{x}_{i}\tau^x_{i+1} &   \forall \: i \in \{ \mathcal{N^+}, \mathcal{E}^- \}\\
 h_x\sum_{i} \tau^{x}_{i}\big( \tau^{x}_{i+\frac{1}{2},+} \tau^{x}_{i+\frac{1}{2}, -}  \big)\tau^x_{i+1}& \forall \: i \in  \mathcal{S^+}. 
\end{cases}
\label{TFIMXYEC}
\end{equation}

While the $e$ charges hop freely in the $\mathcal{N}$ region, in the $\mathcal{S}$ region, in addition to hopping, an $e$ charge creates a pair of $m$ charges on the two plaquettes that share the link $(i, i+1)$. These two plaquette operators are described using $\tau_{i+\frac{1}{2},\pm}$. Furthermore, action of $H^x_f$ on the link $l \in \mathcal{E^+}$ is subtle. Apart from creating a pair of $e$ charges (or hopping an $e$ charge over the link $l$), it flips a pair of dimer ($H_d$) and domino ($D_d$) operators flagging that link (see Appendix \ref{field-dynamics-m-anyon-derivation}). The corresponding action is given by,
\begin{align}
\sigma^{y}_{\text{eq}^{+}} \mapsto \frac{1}{2}\tau^{x}_{v_{1}}\tau^{x}_{v_{2}}&\bigg[\tilde{\mathcal{H}}^{y}_{d_{1}}\tilde{\mathcal{H}}^{y}_{d_{2}} +  \tilde{\mathcal{H}}^{x}_{d_{1}}\tilde{\mathcal{H}}^{y}_{d_{2}}\tau^{x}_{\text{dom}_{1}} + \tilde{\mathcal{H}}^{y}_{d_{1}}\tilde{\mathcal{H}}^{x}_{d_{2}}\tau^{x}_{\text{dom}_{2}} \nonumber\\  &+\tilde{\mathcal{H}}^{x}_{d_{1}}\tilde{\mathcal{H}}^{x}_{d_{2}}\tau^{x}_{\text{dom}_{1}} \tau^{x}_{\text{dom}_{2}}\bigg]. \label{eff-action-sigmax-eq}
\end{align}

Here, $\tilde{\mathcal{H}}^{x/y}$ denotes the unitarily rotated dimer-flip operators, $\tilde{\mathcal{H}}^{x/y} = e^{-i\frac{\pi}{8}\sigma^z}\mathcal{H}^{x/y}e^{i\frac{\pi}{8}\sigma^z}$ acting on the two dimer plaquettes ($d_1, d_2$) attached to the link $l\in \mathcal{E}^+$. Since $J$ is always greater than $h_{x}$, the two domino spins ($\tau_{\text{dom}_{1}}$,
$\tau_{\text{dom}_{2}}$) remain rigidly aligned along the $+z$ direction. This effectively suppresses the last three terms in Eq.~\eqref{eff-action-sigmax-eq}, resulting in $\sigma^{y}_{\text{eq}^{+}}\approx \tau^{x}_{v_{1}}(\tilde{\mathcal{H}}^{y}_{d_{1}}\tilde{\mathcal{H}}^{y}_{d_{2}})\tau^{x}_{v_{2}}$. Consequently, hopping an $e$ anyon over the link $\text{eq}^{+}$ requires only two dimer flips, which are energetically much less expensive than domino spin-flips (or creating $m$ anyons), which have $O(J)$ gaps. This is because there is no energy gap associated to dimer ($H_{d}$) fluctuations in the absence of Zeeman fields. Even in the presence of fields, the dimer gap can at maximum scale as $\sim O(h_{x})$. Hence, the local barrier at the equator encountered by the $e$ particles is comparatively weak relative to that for the $m$ charges, permitting a finite transmission probability across the $\mathcal{E}^+$ region.


In the low-field limit, even though the fluctuations mediated by the degenerate dimer ($H_d$) excitations allow $e$ charges to traverse the $\mathcal{E}^+$ region, the process governed by Eq.~\eqref{TFIMXYEC} leads to eventual confinement of the $e$ charge. This can be seen by noting that any $e$ charge hopping process in $\mathcal{S}^+$ over a bond leads to an energy cost of $2\Delta_{m}\sim 4(J-h_{z})$. This leads to a \textit{string tension} for moving the $e$-anyon on the path $r_e$ inside $\mathcal{S}^{+}$, with the tension growing linearly with the distance traversed. This phenomenon, similar to confinement in gauge theories (see the discussion in Appendix \ref{Z2-gauge-mapping-e-dynamics}) leads to a rapid fall of $T_e$ with the size of the second region.

\subsection{Results for $\theta\neq \pi/2$} \label{res-gen-theta}
We now show that the anyon transmission probabilities increase as we depart from the $\theta=\pi/2$ limit. Obviously, it is expected that the transmission probabilities approach unity as $\theta\rightarrow 0$, since the model reduces to the regular toric code. The dynamics for the $m$ charge at arbitrary $\theta$ is trickier to analyze in a controlled fashion, since the trigonometric functions arising in the definition of the domino operator (see Sec.~\ref{pjunction}) make it challenging to determine the effective spin chain Hamiltonian governing the $m$ anyon dynamics. We leave this problem for future work.

We can, however, make progress at small angles $\delta\theta=\pi/2-\theta$. Here, the effective Hamiltonian can be derived and exhibits a structure almost identical to that obtained for the XYJC model, Eq.~\eqref{mXYJC}, with two key differences: (1) the domino spin ($\tau_{\text{dom}}$) coupling is renormalized, and (2) an Ising-like interaction emerges between the domino ($\tau_{\text{dom}}$) and dimer ($\tau_{0}$) spins, originating from the zero-field energy gap of a single dimer excitation. Since, $D_{d}^{2}(\theta)=2(1+H_{d}\cos{\theta})$, we can write,
\begin{align}
 -J\sum_{d\in \mathcal{E}} D_{d}(\theta) &\equiv -\sqrt{2}J \sum_{d\in \mathcal{E}} \big(1+H_{d}\cos{\theta}\big)^{1/2} \tau^{z}_{\text{dom}}  \nonumber\\
& \approx -\Delta_{D}(\theta) \tau^{z}_{\text{dom}} - \frac{\Delta_{H}(\theta)}{2} \tau^{z}_{0}\tau^{z}_{\text{dom}}.\label{modified-domino-dimer-coupling-theta}
\end{align} 
Here, $\tau^{z}_{0}\equiv H_{d}$ according to our previous convention (see Section.~\ref{m-charge-scattering}). The $\theta$-dependent coupling parameters are: (1) the renormalized domino spin coupling, $\Delta_{D}(\theta) = J\big( \sqrt{1+\cos{\theta}} + \sqrt{1-\cos{\theta}} \big)/\sqrt{2} $ and (2) $\Delta_{H}(\theta)$, which is the energy gap to the dimer-flip excitations, defined in Eq.~\eqref{vortex-gap}. Note that Eq.~\eqref{modified-domino-dimer-coupling-theta} is strictly valid near $\theta=\pi/2$. The effective action of Zeeman field ($\propto h_z$), in terms of $\vec{\tau}$ spins, does not obtain any $\theta$-dependent renormalization (see Appendix \ref{theta-neq-pi2-h-action}). Using the WFM formalism, we calculate the transmission probability ($T_m$) as a function of $\delta\theta=\pi/2-\theta$, for different field strengths, $h_z/J$, and at low quasiparticle incident energy, $E_{\text{inc}}\sim \Delta_m$ (see Fig.~\ref{fig:T-vs-theta-phase-junction}). $T_m$ increases as we move away from XYJC limit $(\theta=\pi/2)$.  
\begin{figure}
    \centering
    \includegraphics[width=0.8\columnwidth]{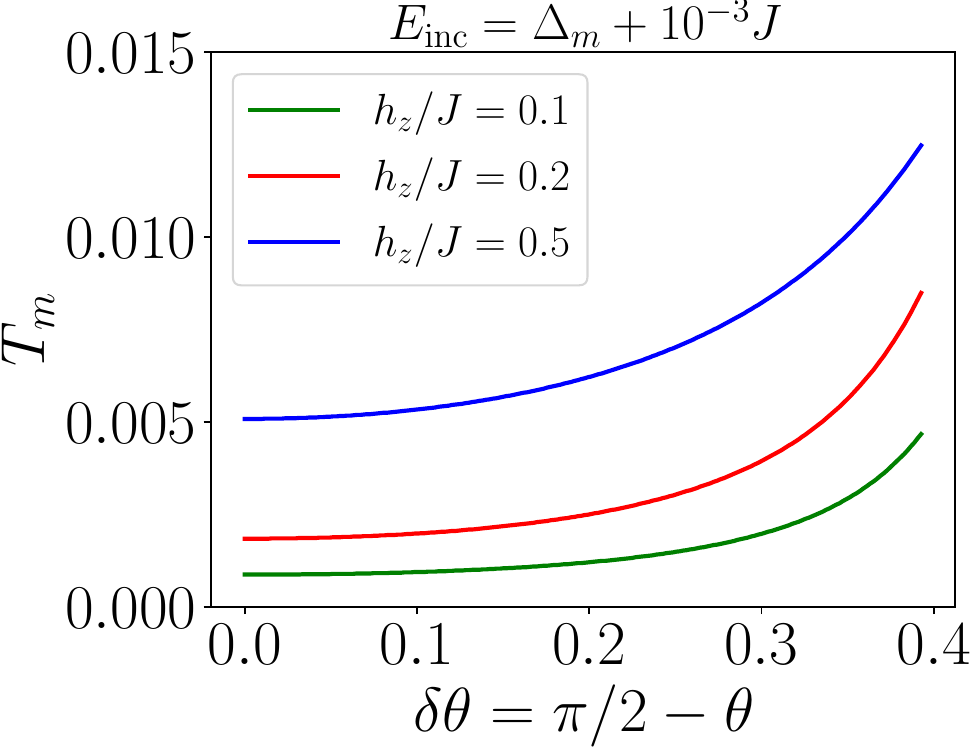}
    \caption{Transmission probability ($T_m$) of $m$ anyons  as a function of junction anisotropy, $\delta\theta=\frac{\pi}{2}-\theta$, for incident energy $E_{inc}\sim \Delta_m =2(J-h_{z})$ and three different values of $h_{z}/J$.}
    \label{fig:T-vs-theta-phase-junction}
\end{figure}

Increase in $T$ with $\delta\theta$ can be understood from the following observations: (1) From Eq.~\eqref{modified-domino-dimer-coupling-theta}, we see that in the absence of $h_z$, the coupling to $\tau^{z}_{\text{dom}}$ modifies to $\Delta_{D}(\theta)-\frac{1}{2}\Delta_{H}(\theta)=\sqrt{2J(1-\cos{\theta})}$ which is $<\sqrt{2J}$ (the corresponding coupling in XYJC). Therefore, when $h_{z}$ is turned on, the domino spin can easily rotate (or flip) from the $+z$-direction. This facilitates $m$ anyon transmission across the junction (also see the discussion at the end of Appendix \ref{continuum-model-m-anyon}). (2) The strength of the Dirac delta barrier also decreases from $\Delta$ (XYJC limit) to $\Delta-\Delta_{H}(\theta)\langle\tau^{z}_{\text{dom}}\rangle$. This further increases $m$ anyon transmission across the junction. In both cases, the dimer excitation gap ($\Delta_{H}(\theta)$) plays the pivotal role. 

We now discuss the dynamics of the $e$ charge. The dynamics of $e$ charges are implemented by $\sigma^x_l$ (along the $r_e$ string), which gets rotated to $\cos{\theta}\sigma^x_l-\sin{\theta}\sigma^y_l$ in the $\mathcal{S}$ region under the unitary rotation Eq.~\eqref{unitary1}. Whereas $\sigma^y$ creates a pair of $m$ charges, in addition to the hopping of $e$ particles, the action of $\sigma^x$ does not cause any fluctuations of $m$ charges. As a result, the $e$ charge can propagate with hopping amplitude $\sim h_x\cos{\theta}$ in the $\mathcal{S}$ region, bypassing the linearly increasing confining potential. Nevertheless, the mismatch in $e$ anyon bandwidth (proportional to its hopping amplitude) between the two regions will give rise to finite junction resistance and the transmission probability, $T_e$ remains less than one for $\theta\neq 0$. 

\section{Conclusion}\label{sec-coclude}

Our investigation is an initial foray to analyze scattering of Abelian anyons in inhomogeneous $\mathbb{Z}_{2}$ spin liquids, for which we consider simple TC variants. As formulated, we have introduced two classes of junctions - one inhomogeneous in the coupling strengths and the other using a set of non-commuting junction operators. These are manifestly semi-permeable and selective, preferentially distinguishing the transport signatures of $e$ and $m$ anyons depending on the initial energy and direction of external field strengths. We have explicitly demonstrated the physical mechanisms behind anyon transmission, which involve effective potential barriers and pseudospin fluctuations. Creating other minimal examples of topologically ordered junctions, analogous to more conventional electronic junctions, either in the context of semiconductors or superconductors, will be exciting. While our work describes $\mathbb{Z}_2$ charges and one-dimensional scattering paths, a generalization to other Abelian/non-Abelian charges and two-dimensional scattering \cite{hart2021correlation, PhysRevB.104.115114, PhysRevB.109.075108, yang2025anyonbraidingpumpprobe} are natural future extensions. 
The latter may even have interesting interferometric signals arising from the braiding of the anyonic charges. Similarly, while our work illustrates the scattering mechanisms in the context of toric-code models, a more general formulation of the same, just using the content of topological order, will be an interesting future direction. Although experimental realizations of TCs are few \cite{toric_code_rydberg_Lukin, satzinger2021realizing, iqbal2024non}, the importance of a concrete experimental proposal of TC junctions for tunable anyonic transport cannot be overemphasized. 


\section{Acknowledgments}

S.B.~acknowledges useful discussions with Nicholas O'Dea, Yaodong Li and Jeongwan Haah. S.B. is supported in part by the US Department of Energy, Office of Basic Energy Sciences, Division of Materials Sciences and Engineering, under Contract No. DE-AC02-76SF00515S. S.S.~acknowledges support from Institute Postdoctoral Fellowship, IIT Kanpur.  Numerical calculations were performed on the workstation $\textit{Sherlock}$ at Stanford University and {\it Syahi} at IITK.


\appendix 
\section{Mapping of the $J_1$-$J_2$ junction in a Zeeman field to an Ising chain}\label{Mapping-J1-J2-model-to-Ising}
As outlined in the main text (see Sec.~\ref{sec-j1j2-map-to-ising}), the plaquette ($B^{x}_{p}$) and star ($A_{v}^{z}$) operators can be equivalently expressed in terms of Ising-like $\vec{\tau}$ operators \cite{PhysRevLett.98.070602, PhysRevB.83.075124, PhysRevB.100.125159}, located at the plaquette centers and vertices of the square lattice, respectively. We next derive the effective actions of the Zeeman terms $-h_{\alpha}\sum_{l}\sigma^{\alpha}_{l}$ in terms of the $\vec{\tau}$ spins. In general, the ``gauge spins'' ($\sigma^{\alpha}_{l}$) cannot be fully eliminated due to the long-range statistical interaction between $e$ and $m$ anyons \cite{Savary_2017}. However, in our setup, the magnetic field acts along distinct one-dimensional chains. Consequently, the mutual statistics between $e$ and $m$ anyons do not influence their behavior, allowing them to act as ordinary bosonic particles. 

To derive the action of $\sigma^x_l$ (which leads to $e$ charge fluctuations), we determine what $\sigma^{x}_l$ does to a general eigenstate of the zero-field toric code, when $m$ charges are totally absent. A general eigenstate in the zero $m$ sector is given by,
\begin{align}
\ket{r^{v}_{1}, r^{v}_{2},...., r^{v}_{N}}= \prod_{i\in v}\big(1+(-1)^{r^{v}_{i}}A_{i}^{z}\big)\ket{\uparrow \uparrow \uparrow...\uparrow}_{x}
\end{align}
where $r_{i}^{v}=0 $ or $1$ for $i\in [1,N]$. The action of $A_{j}^{z}$ and $\sigma^{x}_{l}$ ($l\equiv \langle j,j+1\rangle$) are 
\begin{align}
&A_{j}^{z}\ket{r^{v}_{1}, r^{v}_{2},...., r^{v}_{N}} = (-1)^{r^{v}_{j}} \ket{r^{v}_{1}, r^{v}_{2},...., r^{v}_{N}}\\
&\sigma^{x}_{l}\ket{r^{v}_{1}, r^{v}_{2},...., r^{v}_{N}} =  \ket{r^{v}_{1},..., r^{v}_{j}+1, r^{v}_{j+1}+1,...., r^{v}_{N}}
\end{align}
where addition is defined modulo 2. Therefore,
\begin{align}
\sigma^{x}_{l} \equiv \sigma^{x}_{j,j+1} \mapsto \tau^{x}_{j}\tau^{x}_{j+1} ,
\end{align}
where $\tau^{x}_{j}$ represents the $e$ anyon creation operator at vertex $j$. While our study is in the zero-flux sector, the mapping is valid even in the presence of any (static) background $m$ charges. Here, given that the field is along a specific path, a unitary transformation on a set of $\sigma$s can keep the spins along that path of the reference state ($\ket{\uparrow \uparrow \uparrow...\uparrow}_{x}$) to remain $x$ polarized. 

\indent Similarly, to get the field-induced dynamics for the $m$ anyons, we consider the state  $\ket{r^{p}_{1}, r^{p}_{2},...., r^{p}_{N}}= \prod_{i\in v}\big(1+(-1)^{r^{p}_{i}}B_{i}^{x}\big)\ket{\uparrow \uparrow \uparrow...\uparrow}_{z}$ and see how it changes under $\sigma^{z}_{l}$. We again find $B^{x}_{p}\mapsto \tau^{z}_{p}$ and $\sigma^{z}_{l}\mapsto \tau^{x}_{p}\tau^{x}_{p+1}$, where the dual lattice link $(p,p+1)$ is perpendicular to the (direct lattice) link, $l$. 

\indent Note that the above mapping retains the algebra of the original operators. For instance, $\lbrace B_p^x, \sigma_{l}^z\rbrace=0$ if they share a common link and commute otherwise; this is true for the $\tau$ spins as well since $[\tau^{z}_{p(v)}, \tau^{x}_{p'(v')}\tau^{x}_{p''(v'')}]=0$ if $p(v)\neq p'(v'),p''(v'')$  and anti-commute otherwise. 

\section{Properties of XYJC: Anyon density in the zero-field ground state and topological degeneracy} \label{xyjc-appendix-gs-prop}

\subsection{Anyon density in the ground state}

The ground state manifold of the usual TC is free of anyonic quasiparticles unless we explicitly make some of the star ($A_{v}$) or plaquette ($B_{p}$) couplings negative. In the XYJC model, operator anti-commutativity along the junction results in a non-zero (quantum fluctuating) density of anyons. The ground state is free of $e$ anyons, and $\langle A_{v}^{z}\rangle=+1$ is satisfied everywhere. Also, there are no $m$ anyons in $\mathcal{N}^{+}$, and $\mathcal{S}^{+}$ regions ($\langle B_{p}^{x}\rangle=+1$, for $p\in \mathcal{N}^{+}, \mathcal{S}^{+}$). In the $\mathcal{E}_{q}$ region, $\lbrace B^{x}_{p_{l}^{-}}, B^{yx^{3}}_{p_{l}^{+}}\rbrace=0$; therefore, $m$ anyon density is not necessarily zero here. To see this, express the domino ($D_{d}$) and dimer ($H_{d}$) operator of a single rectangular plaquette in the $\mathcal{E}_{q}$ region, in the two $m$ anyon basis $\ket{B^{x}_{p_{l}^{-}}, B^{x}_{p_{l}^{+}}}$ $=\lbrace \ket{++}, \ket{+-}, \ket{-+}, \ket{--}\rbrace$. Since the dimer operator ($H_{d}=B^{x}_{p_{l}^{-}}B^{x}_{p_{l}^{+}}$) is a symmetry of the XYJC Hamiltonian and commutes with $D_{d}$, the action of the latter preserves the parity of $m$ anyons within $\mathcal{E}_{q}$: $D_{d}$ either acts within $\lbrace \ket{++}, \ket{--}\rbrace$ or $\lbrace \ket{+-}, \ket{-+}\rbrace$. Hence, $D_{d}=D_{d}^{(h_{d}=+1)}\bigoplus D_{d}^{(h_{d}=-1)}$, where  

\begin{align}
 D_{d}^{(h_{d}=+1)}=D_{d}^{(h_{d}=-1)}  =\frac{1}{\sqrt{2}}\begin{pmatrix}
  1 & 1\\
  1 & -1
 \end{pmatrix} 
\end{align}
The eigenvalues $\lambda_{\pm}=\pm 1$ and the normalized eigenstates are: (a) $\ket{D_{d}=+1}= [\cos{(\pi/8)},\ \sin{(\pi/8)}]^{T}$ for $\lambda=+1$, and $\ket{D_{d}=-1}= [\sin{(\pi/8)},\ -\cos{(\pi/8)}]^{T}$ for $\lambda=-1$. Therefore, the degenerate ground state of a single domino plaquette is given by
\begin{align}
\ket{D_{d}=+1}& = c_{1}\bigg(\cos{\frac{\pi}{8}}\ket{++}+\sin{\frac{\pi}{8}}\ket{--}\bigg)\nonumber\\
&\ \ \ \ +c_{2}\bigg(\cos{\frac{\pi}{8}}\ket{+-}+\sin{\frac{\pi}{8}}\ket{-+}\bigg)
\end{align}
Here, $c_{1}$, $c_{2}$ are arbitrary complex numbers, satisfying the normalization condition, $|c_{1}|^{2}+|c_{2}|^{2}=1$. Let us calculate the ground state average of $m$ anyon density at the location of rectangular plaquette $d$,
\begin{align}
n_{d}^{(m)}=\frac{1}{2}\bra{D_{d}=+1} \big(n_{p_{l}^{-}}^{(m)}+n_{p_{l}^{+}}^{(m)}\big)\ket{D_{d}=+1}
\end{align}
where $n^{(m)}_{p_{l}^{\mp}}=(1-B^{x}_{p_{l}^{\mp}})/2$ denotes the $m$ anyon occupation at plaquettes $p_{l}^{\mp}$, located left and right of the equator. A straightforward calculation gives
\begin{align}
n_{d}^{(m)}= \frac{1}{2}-\frac{|c_{1}|^{2}}{2}\cos{\frac{\pi}{4}}    
\end{align}
Since $0\leq |c_{1}|\leq 1$, the average $m$ anyon density ranges between $ \sin^{2}(\pi/8)\leq n_{d}^{(m)}\leq \sin^{2}{(\pi/4)}$, which is finite even without the external Zeeman fields. \\

\subsection{Topological degeneracy}

The ground state (GS) degeneracy discussed so far arises from the local symmetries ($H_{d}$) inherent to our junction model and is unrelated to the topology of the spatial manifold on which the model is defined. As we will elaborate, the topological degeneracy of the phase-junction ground state remains identical to that of the standard toric code.  Consequently, the four distinct topological sectors of excitations ($\mathrm{1}, e, m, \psi$) observed in the canonical toric code are preserved.

Define two toric cycle operators (\textit{Wilson loops}) that act non-trivially on the XYJC ground state, Eq.~\eqref{phys-gs-xytc}, given by
\begin{align}
& W_{x}=\prod_{l\in C_{x}}(\sigma^{x}_{l}\big)^{1-\delta_{l,l_{\text{eq}}}} \bigg(\frac{\sigma^{x}_{l}+\sigma^{y}_{l}}{\sqrt{2}}\bigg)^{\delta_{l,l_{\text{eq}}}} \nonumber\\
& W_{y}=\prod_{l\in C_{y}}(\sigma^{x}_{l}\big)^{1-\delta_{l,l_{\text{eq}}}}\bigg(\frac{\sigma^{x}_{l}+\sigma^{y}_{l}}{\sqrt{2}}\bigg)^{\delta_{l,l_{\text{eq}}}} \label{thooft-loops}
\end{align}
Here, $C_{x,y}$ denote two non-contractible closed loops of the torus that pass through the bonds of the square lattice. For every link, $l\in C_{x,y}$, we apply $\sigma^{x}_{l}$. However, when parts of the loops lie along the equator (i.e., if $l=l_{\mathcal{E}}$ for $l_{\mathcal{E}}\in \mathcal{E}_{q}$), $\sigma^{x}_{l}$ is replaced by the operator inside the brackets of Eq.~\eqref{thooft-loops}. Here, $\delta_{l_1,l_2}$ denotes the Kronecker delta. The fourfold degenerate ground states are given by the states,
\begin{align}
\ket{\mu, \nu}=(W_{x})^{\mu}(W_{y})^{\nu} \bigg| \textrm{GS},\ \lbrace r^{h}_{d}\rbrace_{d\in \mathcal{E}_{q}}\ ,\  (-1)^{\sum_{d}r^{h}_{d}}=1 \bigg\rangle
\end{align}
where $\mu,\nu$ take the values 0 and 1. The other set of loop operators ({\it 't Hooft loops}) that equivalently characterize the above four degenerate states are those of the canonical toric code,
\begin{align}
V_{x}=\prod_{l\in \tilde{C}_{x}}\sigma^{z}_{l}\ \ ,\ \ V_{y}=\prod_{l\in \tilde{C}_{y}}\sigma^{z}_{l}
\end{align}
where $\tilde{C}_{x,y}$ are the contours that pass through the dual square lattice, intersecting the direct lattice links. One can easily verify that $\ket{0,0}\equiv \ket{v_{x}=+1, v_{y}=+1}$, $\ket{0,1}\equiv \ket{v_{x}=-1, v_{y}=+1}$, $\ket{1,0}\equiv \ket{v_{x}=+1, v_{y}=-1}$, and $\ket{1,1}\equiv \ket{v_{x}=-1, v_{y}=-1}$, where $v_{x}$ and $v_{y}$ are the eigenvalues of $V_{x}$ and $V_{y}$ respectively. 

\section{Derivation of effective spin-chain models governing field-induced anyon dynamics in the XY junction}

The Hamiltonian for the XY junction subjected to external magnetic fields is given as follows, 
\begin{align}
H = H_{\textrm{XYJC}} -h_{z}\sum_{l\in r_{m}}\sigma^{z}_{l} -h_{x}\sum_{l\in r_{e}}\sigma^{x}_{l}.
\end{align}
Here, $r_{m}$ and $r_{e}$ are different one dimensional contours, respectively, along which the $\sigma^{z}$ and $\sigma^{x}$ operators act (see Fig.~\ref{fig:xy-toric-code} of the main text). Since anyonic motion is restricted to one dimensional chains, we separate the two dimensional Hamiltonian ($H$) into two parts, $H=H_{0}^{'}[p\not \in r_{m}, v \not \in r_{e}]+H_{\text{ladder}}$, where 
\begin{align}
H_{\text{ladder}}&=-J\sum_{p\in r_{m}} \bigg[B^{x}_{p}\delta (p\in \mathcal{N})+B^{y}_{p}\delta (p\in \mathcal{S})\bigg]\nonumber\\
&\ \ \ \ \ \  -J\sum_{v \in r_{e}} A_{v}^{z}
-h_{z}\sum_{l\in r_{m}} \sigma^{z}_{l}-h_{x}\sum_{l\in r_{e}}\sigma^{x}_{l}
\end{align}
Here, the sum over $p$ corresponds to the plaquettes cut by the $r_{m}$ string, and similarly, the sum over $v$ corresponds to vertices positioned on the $r_{e}$ string. As in the zero-field case, we rotate the spins (from $y$ to $x$) in the $\mathcal{S}$ region,
\begin{align}
U^{\dag}H_{\text{ladder}}U=\tilde{H}_{\text{ladder}}= \tilde{H}^{(m)}+\tilde{H}^{(e)}
\end{align}
where 
\begin{align}
\tilde{H}^{(m)}=&-J\sum\limits_{p \in r_{m}}\bigg[\sum_{p \in \mathcal{N}^{+},\mathcal{S}^{+}}B_{p}^{x} + \sqrt{2}\sum_{p \in \mathcal{E}_{q}} D_{d}\bigg]\nonumber\\
&\ \ \ \ \ \ \ \  \ \ \  \ \ \  \ \ \  \ \ \  \ \ \ \ \ \ \  \ \ \ \ \ \ \ \ \  -h_{z}\sum_{l\in r_{m}} \sigma^{z}_{l} \label{H-tilde-m}\\
\tilde{H}^{(e)}=&-J\sum_{v\in r_{e}}A_{v}^{z}-h_{x}\sum_{l\in r_{e}}\big[\delta (l\in \mathcal{N})\sigma^{x}_{l} - \delta (l\in \mathcal{S})\sigma^{y}_{l}\big] \label{H-tilde-e}
\end{align}

In this section, we derive the effective 1D Hamiltonians in terms of $\vec{\tau}$ spins that describe the dynamics of $m$ and $e$ anyons across the junction. The method is similar to what we have already done for the $J_{1}$-$J_{2}$ model (Sec. \ref{Mapping-J1-J2-model-to-Ising}). The $z$- and $x$-directional magnetic fields are confined to distinct one-dimensional strings, eliminating the need to account for the long-range interaction between $e$ and $m$ anyons arising from their semionic statistics during the derivation of the effective Hamiltonian. 

\subsection{Field-induced dynamics of $m$ anyons}\label{field-induced-dynamics-m-anyon-supp}
We first consider the action of $-h_{z}\sum_{l\in r_{m}}\sigma^{z}_{l}$ on the zero-field XYJC. Since $\sigma^{z}_{l}$ does not cause any fluctuations in the $e$ sector, we consider an arbitrary eigenstate of the XYJC in the $e$ charge-free sector ($A_{v}^{z}=+1\ \forall\  v$) and find out how $\sigma^{z}_{l}$ modifies it. These eigenstates are labeled by the eigenvalues of $B_{p}^{x}$, $D_{d}$ and $H_{d}$,
\begin{widetext}
\begin{align}
&\big| \lbrace r^{h}_{d}\rbrace'_{d\in \mathcal{E}_{q}}, \lbrace r_{d}\rbrace_{d\in \mathcal{E}_{q}},\lbrace r_{p}\rbrace_{p\in (\mathcal{N}^{+},\mathcal{S}^{+})}\big\rangle= \bigg(1+ (-1)^{\sum_{p\in (\mathcal{N}^{+}, \mathcal{S}^{+})}r_{p}}\prod_{d\in \mathcal{E}_{q}}H_{d}\bigg)\nonumber\\
&\ \ \times\prod_{d\in \mathcal{E}_{q}}\big(1+(-1)^{r^{h}_{d}}H_{d}\big)\big(1+(-1)^{r_{d}}D_{d}\big)\prod_{p\in (\mathcal{N}^{+}, \mathcal{S}^{+})}(1+(-1)^{r_{p}}B^{x}_{p})\bigotimes_{l} \ket{\uparrow}^{z}_{l}\label{gen-eig-state-xy-toric-code}
\end{align}
\end{widetext}
This state gives the eigenvalues $H_{d}=(-1)^{r_{d}^{h}}$, $D_{d}=(-1)^{r_{d}}$, and $B_{p}^{x}=(-1)^{r_{p}}$ in their respective domains of action. The first projection operator ensures that we remain in the physical subspace because the $H_{d}$ projection could give rise to state components that change the $m$ charge parity (or the eigenvalue of $\prod_{p\in \mathcal{N}^{+}, \mathcal{S}^{+}}B^{x}_{p}$) and therefore such states must be killed. With open boundary conditions, there are no constraints on $\lbrace r^{h}_{d}\rbrace$, and we take this boundary condition for the subsequent calculations.  

First, we replace all the Ising-like plaquette operators by Pauli-$z$ operators: (1) $B^{x}_{p}\mapsto \tau^{z}_{p}$ (for $p\in \mathcal{N}^{+}, \mathcal{S}^{+}$), (2) $D_{d}\mapsto \tau^{z}_{\text{dom}}$, and (3) $H_{d} \mapsto \mathcal{H}^{z}_{d}$ (for $d\in \mathcal{E}_{q}$). \\

\noindent (A) {\it Action of $\sigma^{z}_{l}$ in the bulk of $\mathcal{N}^{+}$, $\mathcal{S}^{+}$ regions}: In this case, $\sigma^{z}_{l}$ creates $m$ anyons on the two plaquettes that share the common link $l$. Mathematically, this is expressed as
\begin{align}
&\sigma^{z}_{l}\big| \lbrace r^{h}_{d}\rbrace_{d\in \mathcal{E}_{q}}, \lbrace r_{d}\rbrace_{d\in \mathcal{E}_{q}},\lbrace r_{p}\rbrace_{d\in (\mathcal{N}^{+},\mathcal{S}^{+})}\big\rangle \nonumber\\ 
&\ \ \ = \big| \lbrace r^{h}_{d}\rbrace_{d\in \mathcal{E}_{q}}, \lbrace r_{d}\rbrace_{d\in \mathcal{E}_{q}},\lbrace r_{p}\rbrace'_{d\in (\mathcal{N}^{+},\mathcal{S}^{+})}, r_{p}+1, r_{p'}+1 \big\rangle
\end{align}
where the bond $\langle pp'\rangle$ is perpendicular to the edge $l$. The prime symbol above $\lbrace r_{p}\rbrace$ implies that the set excludes the plaquettes $p$ and $p'$. Therefore, in terms of the $\tau$ spins, the action of $\sigma^{z}_{l}$ is given by,
\begin{align}
\sigma^{z}_{l} \mapsto \tau^{x}_{p}\tau^{x}_{p'} \label{sigmaz-to-tau-inside}
\end{align}

\noindent (B) {\it Action of $\sigma^{z}_{l}$ on the boundary between $\mathcal{E}=\mathcal{E}^{-}\cup\mathcal{E}^{+}$ and ($\mathcal{N}^{+}$, $\mathcal{S}^{+}$) regions}: Here, the action of $\sigma^{z}_{l}$ is non-trivial; besides creating one $m$ charge, it also flips one of the $H_{d}$ eigenvalues and converts the $D_{d}$ operator to a new operator $D'_{d}=(B^{x}_{p_{l}^{-}}-B^{x^{3}y}_{p_{l}^{+}})/\sqrt{2}$, which commutes with all other operators except $D_{d}$. In fact, with $D_d$ it anti-commutes---$\lbrace D_{d}, D'_{d}\rbrace=0$. 

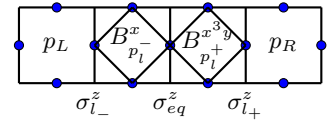
\begin{figure}[htbp]
\centering
\begin{tikzpicture}[scale=1.0]
\foreach \i in {0,1}
{\draw[thick] (0,\i) --(4,\i);}
\foreach \i in {0,1,2,3,4}
{\draw[thick] (\i,0) --(\i,1);}
\foreach \j in {0,1,2,3,4}
{\draw[fill=blue] (\j,0.5) circle [radius=0.06];}
\foreach \i in {0,1}
\foreach \j in {0,1,2,3}
{\draw[fill=blue] (\j+0.5,\i) circle [radius=0.06];}
\draw[thick] (1.5,0.0) --(1.0,0.5); \draw[thick] (1.5,0.0) --(2.0,0.5); \draw[thick] (2.5,0.0) --(2.0,0.5); \draw[thick] (2.5,0.0) --(3.0,0.5);
\draw[thick] (1.5,1.0) --(1.0,0.5); \draw[thick] (1.5,1.0) --(2.0,0.5); \draw[thick] (2.5,1.0) --(2.0,0.5); \draw[thick] (2.5,1.0) --(3.0,0.5);
\node [below, black] at (1.0,0.0) {$\sigma^{z}_{l_{-}}$}; \node [below, black] at (2.0,0.0) {$\sigma^{z}_{eq}$}; \node [below, black] at (3.0,0.0) {$\sigma^{z}_{l_{+}}$};
\node [black] at (1.5,0.5) {$B^{x}_{p_{l}^{-}}$}; 
\node [black] at (0.5,0.5) {$p_{L}$}; 
\node [black] at (2.5,0.5) {$B^{x^{3}y}_{p_{l}^{+}}$};
\node [black] at (3.5,0.5) {$p_{R}$};
\end{tikzpicture}\label{fig-domino-boundary}
\caption{A small slice of the XYJC model, showing the action of $\sigma^{z}_{l_{\pm}}$ and $\sigma^{z}_{\text{eq}}$ in the $\mathcal{E}$ region.}
\end{figure}

For the link $l_{-}$ that resides at the left boundary of the domino ($\mathcal{E}$) region, the explicit expression for the action of $\sigma^{z}_{l_{-}}$ on the general energy eigenstate (Eq.~\eqref{gen-eig-state-xy-toric-code}) of the zero $e$ charge sector is given by,
\begin{widetext}
\begin{align}
&\sigma^{z}_{l_{-}}\big| \lbrace r^{h}_{d}\rbrace_{d\in \mathcal{E}_{q}}, \lbrace r_{d}\rbrace_{d\in \mathcal{E}_{q}},\lbrace r_{p}\rbrace_{d\in (\mathcal{N}^{+},\mathcal{S}^{+})}\big\rangle= \bigg[1+(-1)^{r_{d}^{h}+1}H_{d}\bigg] \prod'_{d'\in \mathcal{E}_{q}}\bigg[1+(-1)^{r_{d'}^{h}}H_{d'}\bigg] 
\nonumber\\
& \times \bigg[1+\frac{(-1)^{r_{d}+1}}{\sqrt{2}}D'_{d}\bigg] \prod'_{d'\in \mathcal{E}_{q}}\bigg[1+\frac{(-1)^{r_{d'}}}{\sqrt{2}}D_{d'}\bigg]\bigg[1+(-1)^{r_{p_{L}}+1}B^{x}_{p_{L}}\bigg] \prod'_{p'\in \mathcal{N}^{+},\mathcal{S}^{+}}\bigg[1+(-1)^{r_{p'}}B^{x}_{p'}\bigg]\bigotimes_{l} \ket{\uparrow}^{z}_{l}\nonumber\\
\end{align}
\end{widetext}
Therefore, $\sigma^{z}_{l}$ flips the $B^{x}_{p_{L}}$ plaquette placed on the left of the link $l_{-}$. It also flips the eigenvalues of $H_{d}$ and $D_{d}$, but simultaneously changes the operator $\sqrt{2}D_{d}$ ($=B^{x}_{p_{l}^{-}}+B^{x^{3}y}_{p_{l}^{+}}$) to $\sqrt{2}D^{'}_{d}$ ($=B^{x}_{p_{l}^{-}}-B^{x^{3}y}_{p_{l}^{+}}$). The prime symbol above the products indicates that the flipped domino index $d$ and the plaquette $p_{L}$ are excluded.

\indent Since $\sigma^{z}_{l}$ acts locally (i.e., it only changes the eigenvalues of the nearby plaquette operators), we use a simplified labeling of these eigenstates. We label $\big| \lbrace r^{h}_{d}\rbrace_{d\in \mathcal{E}_{q}}, \lbrace r_{d}\rbrace_{d\in \mathcal{E}_{q}},\lbrace r_{p}\rbrace_{p\in (\mathcal{N}^{+},\mathcal{S}^{+})}\big\rangle\equiv \ket{r_{d}^{h}, r_{p_{L}}, r_{d}}$. Therefore, the general eigenstate (in the $A_{v}^{z}=1$ sector) is labeled by the eigenvalues of $H_{d}$, $D_{d}$, and $B^{x}_{p_{L}}$ operators that share a common link with $\sigma^{z}_{l_{-}}$. It is obvious that the state $\sigma^{z}_{l_{-}}\ket{r_{p_{L}}, r_{d}^{h}, r_{d}}$ can have non-zero overlaps \textit{only} with the states having $r_{d}^{h}+1$ and $r_{p_{L}}+1$ (mod 2). Hence, the effective actions in these two cases are just \textit{spin-flip} operations described by $\mathcal{H}^{x}_{d}$ and $\tau^{x}_{p_{L}}$. To figure out what happens in the domino ($\lbrace r_{d}\rbrace$) subspace, we have to evaluate the following two overlaps $(A)$ and $(B)$,
\begin{align}
&(A)\ \ \bra{r_{d}^{h}+1, r_{p_{L}}+1, r_{d}}\ \sigma^{z}_{l_{-}}\ \ket{r_{d}^{h}, r_{p_{L}}, r_{d}}\\
&(B)\ \ \bra{r_{d}^{h}+1, r_{p_{L}}+1, r_{d}+1}\ \sigma^{z}_{l_{-}}\ \ket{r_{d}^{h}, r_{p_{L}}, r_{d}}
\end{align}
that measure the extent of flipping of the domino operator. A straightforward calculation yields, 
\begin{align}
&\bra{r_{d}^{h}+1, r_{p_{L}}+1, r_{d}}\ \sigma^{z}_{l_{-}}\ \ket{r_{d}^{h}, r_{p_{L}}, r_{d}} \nonumber\\
&\ \ \ \ \ \ \ \ \ \ = \frac{1}{\sqrt{2}}\exp{\big[-i\frac{\pi}{4}(-1)^{r^{h}_{d}}\big]},\\
&\bra{r_{d}^{h}+1, r_{p_{L}}+1, r_{d}+1}\ \sigma^{z}_{l_{-}}\ \ket{r_{d}^{h}, r_{p_{L}}, r_{d}}\nonumber\\
&\ \ \ \ \ \ \ \ \ = \frac{1}{\sqrt{2}}\exp{\big[i\frac{\pi}{4}(-1)^{r^{h}_{d}}\big]}
\end{align}
Therefore, the effective action of $\sigma^{z}_{l_{-}}$ is the following,
\begin{align}
\sigma^{z}_{l_{-}}\mapsto& \frac{1}{\sqrt{2}} \tau^{x}_{p_{L}}\mathcal{H}^{x}_{d}\big (e^{i\frac{\pi}{4}\mathcal{H}^{z}_{d}}\tau^{x}_{\text{dom}}+e^{-i\frac{\pi}{4}\mathcal{H}^{z}_{d}} \mathds{1}\big)\nonumber\\
&= \frac{1}{2}(1+\tau^{x}_{\text{dom}})\tau^{x}_{p_{L}}\mathcal{H}^{x}_{d}-\frac{1}{2}(1-\tau^{x}_{\text{dom}})\tau^{x}_{p_{L}}\mathcal{H}^{y}_{d} \label{eff-sigmazl}
\end{align}

The steps of obtaining the action of $\sigma^{z}_{l_{+}}$ are almost the same as those done for $\sigma^{z}_{l_{-}}$, except for a change in the signs of the phase factors, because,
\begin{align}
 \sigma^{z}_{l_{+}}\bigg(1+(-1)^{r_{d}}\frac{D}{\sqrt{2}}\bigg)=   \bigg(1+(-1)^{r_{d}}\frac{D'}{\sqrt{2}}\bigg)\sigma^{z}_{l_{+}}
\end{align}
This is in contrast to the case where $\sigma^{z}_{l_{-}}$ is passed through $D_{d}$; there, the operator $\sqrt{2}D_{d}$ ($=B^{x}_{p_{l}^{-}}+B^{x^{3}y}_{p_{l}^{+}}$) converts to $\sqrt{2}D^{'}_{d}$ ($=B^{x}_{p_{l}^{-}}-B^{x^{3}y}_{p_{l}^{+}}$) with an extra minus sign. This extra phase also enters in the overlaps, and as a consequence,
\begin{align}
\sigma^{z}_{l_{+}}&\mapsto \frac{1}{\sqrt{2}} \tau^{x}_{p_{R}}\mathcal{H}^{x}_{d}\big (e^{-i\frac{\pi}{4}\mathcal{H}^{z}_{d}}\tau^{x}_{\text{dom}}+e^{i\frac{\pi}{4}\mathcal{H}^{z}_{d}} \mathds{1}\big)\nonumber\\
&\ \ =\frac{1}{2}(1+\tau^{x}_{\text{dom}})\mathcal{H}^{x}_{d}\tau^{x}_{p_{R}}+\frac{1}{2}(1-\tau^{x}_{\text{dom}})\mathcal{H}^{y}_{d}\tau^{x}_{p_{R}}. \label{eff-sigmazr}
\end{align}
\noindent (C) {\it Action of the $\sigma^{z}_{l}$ operator that lives on the equator}: We call this spin $\sigma_{\textrm{eq}}^z$. In this case, $\lbrace\sigma^{z}_{\textrm{eq}}, D_{d}\rbrace=0$, but commute with the rest of the operators present in the Hamiltonian and with $H_{d}$. Therefore,
\begin{align}
\sigma^{z}_{\text{eq}} \mapsto \tau^{x}_{\text{dom}} \label{eff-sigmaz-eq}
\end{align}

We now express the Hamiltonian (Eq.\eqref{H-tilde-m}) that describes the $m$ charge dynamics, in terms of the dual ($\tau^{\alpha}_{p}$, $\tau^{\alpha}_{\text{dom}}$ and $\mathcal{H}^{\alpha}_{d}$) spins. Using Eqs.~\eqref{sigmaz-to-tau-inside}, \eqref{eff-sigmazl}, \eqref{eff-sigmazr}, and \eqref{eff-sigmaz-eq}, we obtain the following,
\begin{widetext}
\begin{align}
\tilde{H}^{(m)} = & - J\sum_{p\in \mathcal{N}^{+},\mathcal{S}^{+}} \tau^{z}_{p} - h_{z}\sum_{\substack{p\in \mathcal{N}^{+},\mathcal{S}^{+}\\ p\neq p_{L}, p_{l}^{+}}} \tau^{x}_{p}\tau^{x}_{p+1} -\big(\sqrt{2}J \tau^{z}_{\text{dom}} + h_{z} \tau^{x}_{\text{dom}}\big) \nonumber\\
& -\frac{h_{z}}{2}\bigg\lbrace (\tau_{p_L}^x+\tau_{p_R}^x)\mathcal{H}_{d}^x (\tau^x_{\textrm{dom}}+\mathds{1}) +(\tau_{p_L}^x-\tau_{p_R}^x)\mathcal{H}^y_{d}(\tau_{\textrm{dom}}^x-\mathds{1}) \bigg\rbrace
\end{align}
\end{widetext}
The first sum (over plaquettes ($p$)) excludes $p_{l}^{\pm}$ ($\in \mathcal{E}^{\pm}$) and the second sum excludes $p_{L}$ ($\in \mathcal{N}^{+}$) and $p_{l}^{+}$ ($\in \mathcal{E}^{+}$). We rewrite this Hamiltonian as a spin-1/2 quantum Ising chain model interacting with an impurity spin-1/2 positioned at the chain's center, under the influence of a Zeeman field.

\indent The spin chain is constructed from the plaquette centers ($p$) such that $p\in (-\infty, p_{L}] \cup [p_{R}, \infty)$. Then, we make the dimer $\mathcal{H}^{\alpha}_{d}$ part of the chain, placing it between $p_{L}$ and $p_{R}$, and then label it by $\tau^{\alpha}_{0}$. In this way, a $\tau$ spin chain is obtained, which is coupled to the domino spin, $\tau_{\text{dom}}$ (which we referred to as the impurity spin). Therefore, the Hamiltonian becomes Eqs.~\eqref{mXYJC} of the main text.


\subsection{Field-induced dynamics of $e$ anyons} \label{field-dynamics-m-anyon-derivation}
We now consider the action of $-h_{x}\sum_{l\in r_{e}}\sigma^{x}_{l}$ on the zero-field XYJC. After the unitary rotation (of spins), $\sigma^{x}\rightarrow \sigma^{y}$, in the $\mathcal{S}$ region (see Eq.~\eqref{H-tilde-e}). Therefore, the magnetic field also leads to fluctuations in the $m$ sector. We replace $A_{v}^{z} \rightarrow \tau^{z}_{v}$ to denote the presence or absence of an $e$ anyon. In the $\mathcal{N}$ region, $\sigma^{x}_{l}$ creates two $e$ charges at the two ends of the horizontal link $l$ (see Fig.~\ref{fig:e-charge-action}),
\begin{align}
-h_{x}\sum_{l\in r_{e}}\delta(l\in \mathcal{N})\sigma^{x}_{l} = -h_{x}\sum_{\langle vv'\rangle \in  \mathcal{N}} \tau^{x}_{v}\tau^{x}_{v'}\label{eff-action-sigmax-N-region}
\end{align}
where $\langle vv'\rangle\equiv l$ denotes the nearest-neighbor links of the lattice in the $\mathcal{N}$ region that connect two vertices $v$ and $v'=v+1$.
In the southern ($\mathcal{S}^{+}$) region, the $\sigma^{y}$ field creates both $e$ and $m$ charges in pairs (see Fig.~\ref{fig:e-charge-action}),
\begin{align}
-h_{x}\sum_{l\in r_{e}}\delta(l\in \mathcal{S}^{+})\sigma^{y}_{l} = -h_{x}\sum\limits_{\substack{\langle v,v'\rangle\in \mathcal{S}^{+} \\ \langle v,v'\rangle \perp \langle p,p'\rangle}} \tau^{x}_{v} \tau_{v'}^{x}\tau^{x}_{p} \tau_{p'}^{x} \label{eff-action-sigmay-S-region}
\end{align}
The $m$ charges are created on the two plaquettes $p$, $p'$, that share the common link $l$.

\begin{figure}[htbp]
    \centering
    \includegraphics[width=1.0\linewidth]{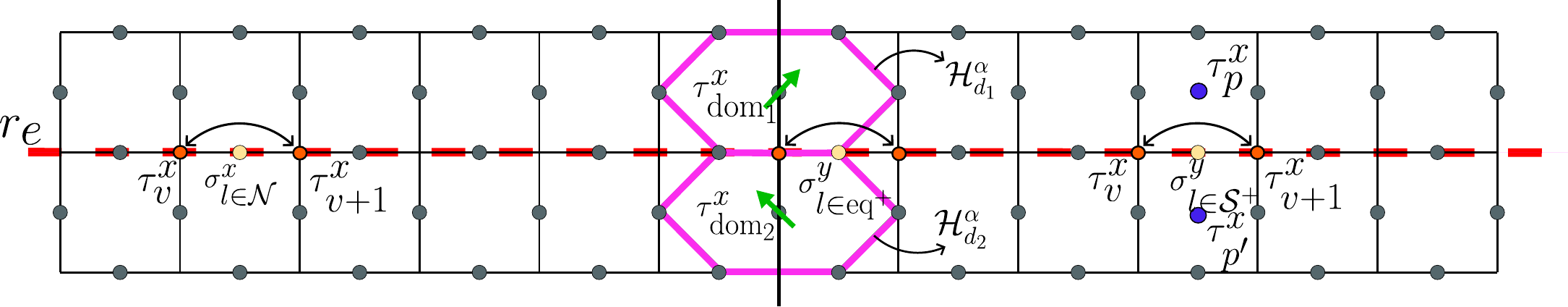}
    \caption{Diagram illustrating various field-driven processes of $e$ charge dynamics along the $r_{e}$ string in the dual spin ($\tau$) framework.}
    \label{fig:e-charge-action}
\end{figure}

For the link $l=\text{eq}^{+}$, located directly to the right of the junction, the action of $\sigma^{y}_{\text{eq}^{+}}$ has several consequences (see Fig.~\ref{fig:e-charge-action}), in addition to creating two $e$ charges at its two ends. (i) The eigenvalues of the two {\it dimer} operators ($H_{d_{1}},\ H_{d_{2}}$) that share the common link $\text{eq}^{+}$, are flipped. (ii) The two {\it domino} operators $D_{d_{1}}$, and $D_{d_{2}}$, which also share the link $\text{eq}^{+}$ will transform to $D'_{d_{1}}$, and $D'_{d_{2}}$ respectively. Hence, to evaluate the effective action of $\sigma^{y}_{\text{eq}^{+}}$, we must follow a similar procedure used to determine the action of $\sigma^{z}$ across the junction (see the discussion in Sec.~\ref{field-induced-dynamics-m-anyon-supp}).

\indent Consider the general eigenstate of the zero-field XYJC model in the $z$ basis (Eq.~\eqref{gen-eig-state-xy-toric-code}) that has no $e$ charges. Since $\sigma^{y}_{l}$ creates two $e$ charges at the ends of the link $l$, we create such a state from Eq.~\eqref{gen-eig-state-xy-toric-code}, by flipping the spin at the $\text{eq}^{+}$ link of the reference state, 
\begin{align}
&\big| \lbrace r^{h}_{d}\rbrace_{d\in \mathcal{E}_{q}}, \lbrace r_{d}\rbrace_{d\in \mathcal{E}_{q}},\lbrace r_{p}\rbrace_{p\in (\mathcal{N}^{+},\mathcal{S}^{+})}, r^v_{1}, r^v_{2}\big\rangle \nonumber \\
&=\prod_{d\in \mathcal{E}_{q}}\big[1+(-1)^{r^{h}_{d}}H_{d}\big]\bigg[1+(-1)^{r_{d}}D_{d}\bigg]\nonumber\\
&\ \ \times\prod_{p\in \mathcal{N}^{+}, \mathcal{S}^{+}}[1+(-1)^{r_{p}}B^{x}_{p}]\bigg\lbrace \ket{\downarrow}_{\text{eq}^{+}}\bigotimes_{l\not \in \text{eq}^{+}} \ket{\uparrow}^{z}_{l}\bigg\rbrace
\end{align}
The (static) $e$ anyons are placed in this state on the two nearest-neighbor vertices $v_{1}$ and $v_{2}$ connected by $\text{eq}^{+}$.
We consider the action of $\sigma^{y}_{\text{eq}^{+}}$ on the eigenstate Eq.~\eqref{gen-eig-state-xy-toric-code},  
\begin{widetext}
\begin{align}
&\sigma^{y}_{\text{eq}^{+}}\big| \lbrace r^{h}_{d}\rbrace_{d\in \mathcal{E}_{q}}, \lbrace r_{d}\rbrace_{d\in \mathcal{E}_{q}},\lbrace r_{p}\rbrace_{p\in (\mathcal{N}^{+},\mathcal{S}^{+})}\big\rangle= \prod_{i=1,2}\big(1+(-1)^{r^{h}_{d_{i}}+1}H_{d_{i}}\big)\prod_{\substack{d\in \mathcal{E}_{q} \\ d\neq d_{1}, d_{2}}}\big(1+(-1)^{r^{h}_{d}}H_{d}\big) \nonumber\\
&\ \ \ \ \ \ \times \prod_{i=1,2}\big(1+(-1)^{r_{d_{i}}}D'_{d_{i}}\big)\prod_{\substack{d\in \mathcal{E}_{q} \\ d\neq d_{1}, d_{2}}}\big(1+(-1)^{r_{d}}D_{d}\big)
 \prod_{p\in \mathcal{N}^{+}, \mathcal{S}^{+}}\bigg(1+(-1)^{r_{p}}B^{x}_{p}\bigg)\bigg\lbrace \ket{\downarrow}_{\text{eq}^{+}}\bigotimes_{e\not \in \text{eq}^{+}} \ket{\uparrow}^{z}_{e}\bigg\rbrace
\end{align}
\end{widetext}
To simplify the notation, we denote $\big| \lbrace r^{h}_{d}\rbrace_{d\in \mathcal{E}_{q}}, \lbrace r_{d}\rbrace_{d\in \mathcal{E}_{q}},\lbrace r_{p}\rbrace_{p\in (\mathcal{N}^{+},\mathcal{S}^{+})}\big\rangle$ as $\ket{r_{d_{1}}, r_{d_{2}}, r_{d_{1}}^{h}, r^{h}_{d_{2}}}$ and $\big| \lbrace r^{h}_{d}\rbrace_{d\in \mathcal{E}_{q}}, \lbrace r_{d}\rbrace_{d\in \mathcal{E}_{q}},$ $\lbrace r_{p}\rbrace_{p\in (\mathcal{N}^{+},\mathcal{S}^{+})}, r^v_{1}, r^v_{2}\big\rangle$ as $\ket{r_{d_{1}}, r_{d_{2}}, r_{d_{1}}^{h}, r^{h}_{d_{2}}, r^v_{1}, r^v_{2}}$. We must have, 
\begin{align}
&\sigma^{y}_{\text{eq}^{+}}\ket{r_{d_{1}}, r_{d_{2}}, r_{d_{1}}^{h}, r^{h}_{d_{2}}} \nonumber\\
&=\sum_{p\in \mathcal{R}_{D}} c_{p}\ket{r_{d_{1}}^{h}+1, r^{h}_{d_{2}}+1, r^v_{1}, r^v_{2}}\otimes \ket{p}
\end{align}
where the sum over $p$ runs over the set: $\mathcal{R}_{D}=\big\lbrace (r_{d_{1}}, r_{d_{2}}), (r_{d_{1}}+1, r_{d_{2}}), (r_{d_{1}}, r_{d_{2}}+1), (r_{d_{1}}+1, r_{d_{2}}+1)\big\rbrace$ and $c_p$ are appropriate coefficients. A straightforward calculation yields the following, 
\begin{align}
&c_{(r_{d_{1}}, r_{d_{2}})} = \prod_{i=1,2}\frac{1}{\sqrt{2}}\exp{\left[\textrm{i}\frac{\pi}{4}(-1)^{r_{d_{i}}^{h}}\right]} \nonumber\\
&c_{(r_{d_{1}}+1, r_{d_{2}})}= \frac{1}{2}\exp{\left[-\textrm{i}\frac{\pi}{4}(-1)^{r_{d_{1}}^{h}}\right]}\exp{\left[\textrm{i}\frac{\pi}{4}(-1)^{r_{d_{2}}^{h}}\right]}\nonumber\\
&c_{(r_{d_{1}}, r_{d_{2}}+1)}= \frac{1}{2}\exp{\left[\textrm{i}\frac{\pi}{4}(-1)^{r_{d_{1}}^{h}}\right]}\exp{\left[-\textrm{i}\frac{\pi}{4}(-1)^{r_{d_{2}}^{h}}\right]}\nonumber\\
& c_{(r_{d_{1}}+1, r_{d_{2}}+1)}=\prod_{i=1,2}\frac{1}{\sqrt{2}}\exp{\left[-\textrm{i}\frac{\pi}{4}(-1)^{r_{d_{i}}^{h}}\right]}
\end{align}
Hence, the action of $\sigma^{y}_{\text{eq}^{+}}$ is given by,
\begin{widetext}
\begin{align}
\sigma^{y}_{\text{eq}^{+}} & \mapsto \frac{1}{2}\tau^{x}_{v_{1}}\tau^{x}_{v_{2}}\mathcal{H}^{x}_{d_{1}}\mathcal{H}_{d_{2}}^{x}\bigg[ e^{i\frac{\pi}{4}\mathcal{H}^{z}_{d_{1}}} e^{i\frac{\pi}{4}\mathcal{H}^{z}_{d_{2}}} \mathds{1}_{\text{dom}_{1}} \otimes \mathds{1}_{\text{dom}_{2}} + e^{-i\frac{\pi}{4}\mathcal{H}^{z}_{d_{1}}} e^{i\frac{\pi}{4}\mathcal{H}^{z}_{d_{2}}} \tau^{x}_{\text{dom}_{1}}\otimes \mathds{1}_{\text{dom}_{2}}\nonumber\\
&\ \ \ \  + e^{i\frac{\pi}{4}\mathcal{H}^{z}_{d_{1}}} e^{-i\frac{\pi}{4}\mathcal{H}^{z}_{d_{2}}} \mathds{1}_{\text{dom}_{1}}\otimes \tau^{x}_{\text{dom}_{2}} + e^{-i\frac{\pi}{4}\mathcal{H}^{z}_{d_{1}}} e^{-i\frac{\pi}{4}\mathcal{H}^{z}_{d_{2}}} \tau^{x}_{\text{dom}_{1}}\otimes \tau^{x}_{\text{dom}_{2}}\bigg]
\end{align}
\end{widetext}
Using the identity, $\mathcal{H}^{x}e^{\pm i\pi \mathcal{H}^{z}/4} = (\mathcal{H}^{x}\pm \mathcal{H}^{y})/\sqrt{2}$, we can simplify the above as follows. We work in a rotated basis for the $\mathcal{H}^{\alpha}$ spins, $\tilde{\mathcal{H}}^{x} = (\mathcal{H}^{x}-\mathcal{H}^{y})/\sqrt{2}$, $\tilde{\mathcal{H}}^{y} =(\mathcal{H}^{x}+\mathcal{H}^{y})/\sqrt{2}$, $\tilde{\mathcal{H}}^{z} =\mathcal{H}^{z}$. The action of $\tilde{\mathcal{H}}^{x,y}$ on $\mathcal{H}^{z}$ eigenstates are: $\tilde{\mathcal{H}}^{x}\ket{\mathcal{H}^{z}=\pm 1}= e^{\mp i\pi/4}\ket{\mathcal{H}^{z}=\mp 1}$, and $\tilde{\mathcal{H}}^{y}\ket{\mathcal{H}^{z}=\pm 1}= e^{\pm i\pi/4}\ket{\mathcal{H}^{z}=\mp 1}$. Thus, up to a phase factor, the actions of $\tilde{\mathcal{H}}^{x,y}$ are basically {\it spin-flip} operations. In terms of these new spin variables, we obtain Eq.\eqref{eff-action-sigmax-eq} in the main text. The Eqs.~\eqref{eff-action-sigmax-N-region}, \eqref{eff-action-sigmay-S-region}, and \eqref{eff-action-sigmax-eq} describe the full action of the $\sigma^{x}$ magnetic field for the $e$ charge dynamics.



\section{Jordan-Wigner fermionization of the spin chain model describing $m$ anyon dynamics in XY toric code junction}
The spin chain Hamiltonian (Eq.~\eqref{mXYJC}), which describes the dynamics of $m$ anyons along the chain $r_{m}$, can be mapped to a fermionic chain featuring both hopping and $p$-wave pairing terms, coupled to a spin-$\frac{1}{2}$ impurity at the chain's center. This mapping is achieved through the Jordan-Wigner transformation \cite{sachdev_QPT}, defined as follows:
\begin{align}
&\tau^{x}_{i} =  (f_{i}+f_{i}^{\dag})\prod_{j<i}(1-2f^{\dag}_{j}f_{j}) \nonumber\\
&\tau^{y}_{i} = i(f^{\dag}_{i}-f_{i})\prod_{j<i}(1-2f^{\dag}_{j}f_{j})\nonumber\\
& \tau^{z}_{i} = 1-2f^{\dag}_{i}f_{i}
\end{align}

The spin Hamiltonian (Eq.~\eqref{mXYJC}) transforms to, 

\begin{align}
\tilde{H}^{(m)} = H_{m} +H_{\text{dom}} + H_{I}^{(1)} + H_{I}^{(2)}
\end{align}
where
\begin{align}
H_{m}=-\frac{1}{2}\sum_{i\notin \lbrace -1,0\rbrace}\big[\Psi^{\dag}_{i}(t\tau^{z}+\textrm{i}\Delta\tau^{y})\Psi_{i+1}+\textrm{H.c.}\big] \nonumber\\
-\frac{\mu}{2}\sum_{i\neq 0} \Psi^{\dag}_{i}\tau^{z}\Psi_{i}
\end{align}
\begin{align}
& H_{\text{dom}}=-\big(\sqrt{2}J\tau^{z}_{\text{dom}}+h_z \tau^{x}_{\text{dom}}\big)\\
& H_{I}^{(1)}=-h_{z}\big(\Psi^{\dag}_{-1} \mathbf{\tilde{B}}_{L,1}^{\dag} \Psi_{0} + \Psi^{\dag}_{0}\mathbf{\tilde{B}}_{R,1}^{\dag} \Psi_{1} + \textrm{H.c.}\big) \\
& H^{(2)}_{I}=-h_{z}\big(\Psi^{\dag}_{-1} \mathbf{\tilde{B}}_{L,2}^{\dag} \Psi_{0} + \Psi^{\dag}_{0}\mathbf{\tilde{B}}_{R,2}^{\dag} \Psi_{1} + \textrm{H.c.})\tau^{x}_{\text{dom}}\label{Hm-ladder-fermion-version}
\end{align}
Here, we have used the spinor operators $\Psi_{i}=(f_{i}\ ,f^{\dag}_{i})^{T}$ in the particle-hole space, to write the Hamiltonian in a compact form. Here, $\mu=-2J$, and $t=\Delta=h_{z}$. The $\tilde{B}_{L/R, 1/2}$ matrices are defined as follows,
\begin{align}
\mathbf{\tilde{B}}_{L,1} = -\frac{1}{4}
\begin{bmatrix}
1-\textrm{i} & -(1-\textrm{i}) \\
 1+\textrm{i} & -(1+\textrm{i})    
\end{bmatrix},\mathbf{\tilde{B}}_{R,1} = -\frac{1}{4}
\begin{bmatrix}
1-\textrm{i} & -(1+\textrm{i}) \\
 1-\textrm{i} & -(1+\textrm{i})    
\end{bmatrix}
\end{align}
and, 
\begin{align}
\mathbf{\tilde{B}}_{L,2} = -\frac{1}{4}
\begin{bmatrix}
1+i & -(1+i) \\
 1-i & -(1-i)    
\end{bmatrix},\mathbf{\tilde{B}}_{R,2} = -\frac{1}{4}
\begin{bmatrix}
1+i & -(1-i) \\
 1+i & -(1-i)    
\end{bmatrix}
\end{align}
This form of the Hamiltonian is amenable to a transmission probability computation using the wave-function matching formalism described in the next section. 

\section{Wavefunction matching formalism for calculating transmission probability}\label{wavefunction-matching-transmission}
Here, we describe the framework of the ``wavefunction matching'' approach \cite{kelly_transmission_calculation} to compute $m$ anyon transmission probability for the XY junction problem, governed by the Hamiltonian Eq.~\eqref{Hm-ladder-fermion-version}. The same method is also used for the $J_{1}$-$J_{2}$ junction problem. The key principle is the continuity of wavefunctions, which connects the states far from the scattering center (junction) to those near the scattering region. Leveraging this property, the infinite-dimensional time-independent Schrödinger equation for the entire system can be reduced to solving a finite set of linear inhomogeneous equations that capture the dynamics near the scattering center. 

\indent We first derive the ideal lead eigenstates, i.e., solutions in the absence of impurity or scatterer.

\subsection{Ideal lead solutions}

The Hamiltonian away from the impurity is simply the fermionic Kitaev chain model \cite{kitaev_chain}. In the particle-hole space, the Hamiltonian matrix elements are given by,
\begin{align}
H_{i,i+1} = &\bra{i} H \ket{i+1} =  \mathbf{B}^{\dag}_{L/R} = -\big(t\tau^{z}+i\Delta\tau^{y}\big)/2,\nonumber\\
& H_{i,i} = \bra{i} H \ket{i} = \mathbf{H}_{L/R}= -\mu\tau^{z}/2
\end{align}
where $\mu = -2J,\ t=h,\ \Delta = h$ are the coupling parameters of the left ($L$) and right ($R$) leads. The left and right lead Hamiltonians do not couple to the domino spin ($\tau^{\alpha}_{\text{dom}}$) and therefore act as identity in the domino spin space. 

\indent The energy eigenvalue equation for the leads is given by,
\begin{align}
-\mathbf{H}_{i,i-1} \Phi_{i-1} + (E-\mathbf{H}_{i,i})\Phi_{i} -\mathbf{H}_{i,i+1}\Phi_{i+1}=0  \label{TISE-1}
\end{align}
where $\mathbf{H}_{i,j}$ are the Hamiltonian blocks connecting the sites $i$ and $j$. At every site, there is a 4 dimensional state space, which is a direct product of domino spin (dim=2) and particle-hole (dim=2) spaces,
\begin{align}
\mathbf{H}_{i,i-1} = &\mathds{1}_{2\times 2} \otimes (\mathbf{B}_{L/R})_{2\times 2}\ ,\mathbf{H}_{i,i+1} = \mathds{1}_{2\times 2} \otimes (\mathbf{B}^{\dag}_{L/R})_{2\times 2} \label{H-blocks-lead}\nonumber\\
&\mathbf{H}_{i,i} = \mathds{1}_{2\times 2} \otimes(\mathbf{H}_{L/R})_{2\times 2}
\end{align}
Because of translation symmetry of the leads, the onsite and off-diagonal Hamiltonians are position-independent.

\indent The lead modes have the following structure,
\begin{align}
\Phi_{i} = 
\begin{pmatrix}
c_{+}\\
c_{-}
\end{pmatrix}
\otimes \Psi_{i}
\end{align}
where $c_{\pm}$ are arbitrary complex numbers representing the two spin components of the lead wavefunction. Since there is no coupling between the fermions and the domino spin inside the leads, the time-independent Schrodinger equation (Eq.~\eqref{TISE-1}) can be separated into spin and fermion parts, where the spin part is just the trivial identity operation and for the fermion part, one obtains the following
\begin{align}
-\mathbf{B}_{L/R} \Psi_{i-1} + (E-\mathbf{H}_{L/R})\Psi_{i} -\mathbf{B}^{\dag}_{L/R}\Psi_{i+1}=0  \label{TISE-2}
\end{align}
According to Bloch's theorem, the wave functions at successive sites are related by a phase, and for a two-component wavefunction, it is a (unimodular) matrix, i.e., $\Psi_{i}= F(\pm) \Psi_{i-1}$. Generalization of the usual Bloch phases ($e^{\pm ika}$) to the multi-orbital case ($n_{\textrm{orb}}=2$) leads to the following Bloch matrix \cite{kelly_transmission_calculation}, 
\begin{align}
\mathbf{F}(\pm) = \sum_{n=1}^{2} \lambda_{n,\pm} \mathbf{u}_{n}(\pm)\mathbf{u}^{\dag}_{n}(\pm)\label{block-matrix}
\end{align}  
Here, $\lambda_{n,\pm} =e^{\pm \textrm{i}ka} $ (independent of $n$), where $n$ indexes the orbital. The value of $k$ is related to energy $E$ by the following identity,
\begin{align}
\cos{k} = \frac{E^{2}-(\mu/2)^{2}-t^{2}}{2(\mu/2)t}
\end{align}
The two column vectors $\mathbf{u}_{1}(\pm) = [\cos{(\theta_{k}/2)}$, $
\pm \textrm{i} \sin{(\theta_{k}/2)} ]^{\textrm{T}}$ and $\mathbf{u}_{2}(\pm) = [\pm \textrm{i} \sin{(\theta_{k}/2)},\ \cos{(\theta_{k}/2)}]^{\textrm{T}}$ correspond to the particle- and hole-like normal modes of the time-independent Schrödinger equation, Eq.\eqref{TISE-2}. The $\pm$ denotes the chirality of these states (right or left moving nature). The Bloch matrix is determined from explicit evaluation of Eq.~\eqref{block-matrix}, and is proportional to identity,
\begin{align}
 \mathbf{F}(\pm) =e^{\pm \textrm{i}ka}\mathds{1}_{2\times 2}   
\end{align}
The general (propagating wave) solution of the Bloch equation is given by,
\begin{align}
\Psi_{i} = \mathbf{F}^{i-j}(+)\Psi_{j}(+)+ \mathbf{F}^{i-j}(-)\Psi_{j}(-) \label{Gen-sol}
\end{align}
where $\mathbf{F}^{m}(\pm) = \sum_{n=1}^{2} (\lambda_{n,\pm})^{m}\  \mathbf{u}_{n}(\pm)\mathbf{u}^{\dag}_{n}(\pm) =e^{\pm \textrm{i}mka}\mathds{1}_{2\times 2} $, for any integer $m$.

\subsection{Wavefunction matching}

By matching the states near the scatterer with the traveling Bloch wave solutions, we can derive equations for $\Psi_{-1}$ and $\Psi_{S+2}$ (wavefunction amplitudes near the scattering region), where $S=$ the size of the scattering region. This helps us to truncate the infinite-dimensional Schrodinger equation (of the full problem) to a $(S+2)$ dimensional matrix equation. For the Hamiltonian Eq.~\eqref{Hm-ladder-fermion-version}, $S=1$ (for the single impurity), while for the $J_{1}$-$J_{2}$ model, $S=W$ (the width of the $J_{2}$ region). The wavefunction amplitude at $i=-1$ (which is within the left lead) can be obtained from the general solution, Eq.~\eqref{Gen-sol},
\begin{align}
 \Psi_{-1} = \mathbf{F}_{L}^{-1}(+)\Psi_{0}(+) + \mathbf{F}_{L}^{-1}(-)\Psi_{0}(-)
 \end{align}
We substitute $\Psi_{0}(-) = \Psi_{0}- \Psi_{0}(+)$ (also obtained from Eq.~\eqref{Gen-sol}) in the above equation to obtain,
 \begin{align}
 \Psi_{-1} = \big(\mathbf{F}_{L}^{-1}(+)-\mathbf{F}_{L}^{-1}(-)\big)\Psi_{0}(+) + \mathbf{F}_{L}^{-1}(-)\Psi_{0} \label{psi-1}
 \end{align}
The time-independent Schrodinger equation (Eq.~\eqref{TISE-1}) at $i=0$, which is at the boundary between the scattering region ($i=1,..., S$) and the left lead ($i=-\infty,...,-1,0$), is given by,
\begin{align}
-\mathbf{H}_{0,-1}\Phi_{-1} + (E- \mathbf{H}_{0,0})\Phi_{0} -\mathbf{H} _{0,1}\Phi_{1}=0 \label{TISE-i0}
\end{align}
From Eq.~\eqref{H-blocks-lead}, $\mathbf{H}_{0,-1} = \mathds{1}_{s} \otimes (\mathbf{B}_{L})$, and $\mathbf{H}_{0,0} = \mathbf{H}_{L}$. On the other hand, $\mathbf{H}_{0,1} = \mathds{1}\otimes \mathbf{\tilde{B}}^{\dag}_{L,1}+\tau^{x}\otimes \mathbf{\tilde{B}}^{\dag}_{L,2} $ (hopping matrices connecting the left lead and the site $i=1$ (the impurity site)). Replace $\Psi_{-1}$ in Eq.~\eqref{TISE-i0} by Eq.~\eqref{psi-1}, to obtain the following, 
\begin{align}
&\big[\mathds{1}\otimes \big(E-\mathbf{H}_{L}-\mathbf{B}_{L}\mathbf{F}_{L}^{-1}(-)\big)\big]\Phi_{0} - \big[\mathds{1}\otimes \mathbf{\tilde{B}}^{\dag}_{L,1} \nonumber\\ &+\tau^{x}\otimes \mathbf{\tilde{B}}^{\dag}_{L,2} \big]\Phi_{1}=\big[\mathds{1}\otimes \mathbf{B}_{L}\big(\mathbf{F}_{L}^{-1}(+)-\mathbf{F}_{L}^{-1}(-)\big)\big]\Phi_{0}(+) \label{Eq1-scattering-region}
\end{align} 
where $\Phi_{0}(+)$ is the amplitude at site $i=0$ of the incoming (right-moving) wave coming from the left lead,
\begin{align}
\Phi_{0}(+) = \mathbf{u}_{\sigma}^{s} \otimes \mathbf{u}^{f}_{L,m}(+)\ ;\ \ \ \sigma=\uparrow, \downarrow, m=1,2 
\end{align}
and $\mathbf{u}_{\sigma}^{s} = \lbrace (1,\ 0)^{\textrm{T}},\ (0,\ 1)^{\textrm{T}}\rbrace$ are the two possible eigenstates of $\tau^{z}_{\text{dom}}$. The states $\mathbf{u}^{f}_{L,m}(+)$ are the right-moving eigenmodes of the left lead.

For $i=S+1$, which is at the boundary between the scattering region ($i=1,..., S$) and the right lead ($i=(S+1),..., +\infty$), the time-independent Schrodinger equation is given by,
\begin{align}
-\big[\mathds{1}\otimes\mathbf{\tilde{B}}_{R,1} + & \tau^{x}\otimes \mathbf{\tilde{B}}_{R,2} \big]\Phi_{S} + \big[\mathds{1}\otimes(E-\mathbf{H}_{R})\big] \Phi_{S+1} \nonumber\\
& - \big[\mathds{1}\otimes \mathbf{B}_{R}^{\dag}]\Phi_{S+2} = 0
\end{align}
There is no left-moving wave from the right lead; $\Psi_{S+1}(-)=0$. Therefore, $\Psi_{S+2} = \mathbf{F}_{R}(+)\Psi_{S+1}$. Substituting this in the Schrodinger equation, one obtains,
\begin{align}
-\big[\mathds{1}\otimes\mathbf{\tilde{B}}_{R,1}+ \tau^{x}\otimes \mathbf{\tilde{B}}_{R,2}\big]\Phi_{S} + &\big[\mathds{1}\otimes(E-\mathbf{H}_{R} \nonumber\\
&-\mathbf{B}_{R}^{\dag}\mathbf{F}_{R}(+))\big] \Phi_{S+1} = 0 \label{Eq2-scattering-region}
\end{align}
The time-independent Schrodinger equation at the scattering center ($i=S=1$) is given by,
\begin{align}
-\mathbf{H}_{1,0} \Phi_{0} + (E-\mathbf{H}_{1,1})\Phi -\mathbf{H}_{1,2}\Phi_{2} =0 \label{Eq3-scattering-region}
\end{align}
Here, $\mathbf{H}_{1,0} = \mathds{1}\otimes \mathbf{\tilde{B}}_{L,1}+\tau^{x}\otimes \mathbf{\tilde{B}}_{L,2}$, $\mathbf{H}_{1,2} = \mathds{1}\otimes \mathbf{\tilde{B}}^{\dag}_{R,1}+\tau^{x}\otimes \mathbf{\tilde{B}}^{\dag}_{R,2}$, and $\mathbf{H}_{1,1} = \mathbf{H}_{0} $ is the impurity Hamiltonian, which is given by $\textbf{H}_{0} = - \big(\sqrt{2}J\tau^{z}+h_z \tau^{x}\big) \otimes \mathds{1}_{f}. $


The lead self-energies are given by the following expressions,
\begin{align} \label{lead-self-energy}
&\boldsymbol{\Sigma}_{L}=\mathbf{B}_{L} \mathbf{F}^{-1}_{L}(-) = -\frac{t}{2}e^{ik}(\tau^{z}-i\tau^{y}),\\
&\boldsymbol{\Sigma}_{R}=\mathbf{B}^{\dag}_{R} \mathbf{F}_{R}(+) = -\frac{t}{2}e^{ik}(\tau^{z}+i\tau^{y})
\end{align}    
The Eqs.~\eqref{Eq1-scattering-region}, \eqref{Eq2-scattering-region}, \eqref{Eq3-scattering-region} can be combined to a compact matrix form, 
\begin{widetext}
\begin{align}
E-\mathbf{H}'=
\begin{bmatrix}
\mathds{1}_{s}\otimes \big(E-\mathbf{H}_{L}-\boldsymbol{\Sigma}_{L}\big) & - \big(\mathds{1}_{s}\otimes \mathbf{\tilde{B}}^{\dag}_{L,1}+\tau^{x}\otimes \mathbf{\tilde{B}}^{\dag}_{L,2}\big) & 0\\[2pt]
-\big(\mathds{1}_{s}\otimes \mathbf{\tilde{B}}_{L,1}+\tau^{x}\otimes \mathbf{\tilde{B}}_{L,2}\big)  &  \big(E+ \sqrt{2}J\tau^{z}+h_z \tau^{x}\big) \otimes \mathds{1}_{f} & -\big(\mathds{1}_{s}\otimes \mathbf{\tilde{B}}^{\dag}_{R,1}+\tau^{x}\otimes \mathbf{\tilde{B}}^{\dag}_{R,2}\big) \\[2pt]
0 & -\big(\mathds{1}\otimes\mathbf{\tilde{B}}_{R,1}+ \tau^{x}\otimes \mathbf{\tilde{B}}_{R,2}\big) & \mathds{1}_{s}\otimes \big(E-\mathbf{H}_{R}-\boldsymbol{\Sigma}_{R}\big)
\end{bmatrix}
\end{align}
\end{widetext}
and, 
\begin{align}
(E-\mathbf{H}') 
\begin{bmatrix}
\Phi_{0}\\
\Phi_{1}\\
\Phi_{2}
\end{bmatrix} = 
\begin{bmatrix}
\mathbf{Q}_{0} \\
 0\\
 0
\end{bmatrix},
\end{align}
where
\begin{align}
\mathbf{Q}_{0}= \big(\mathds{1}_{s}\otimes \mathbf{B}_{L}\big[ \mathbf{F}^{-1}_{L}(+)-\mathbf{F}^{-1}_{L}(-)\big]\big)\Phi_{0}(+).
\end{align}
For the $J_{1}$-$J_{2}$ model, there is no impurity (domino) spin. The block matrix $\textbf{H}'$ has size $2(S+2)\times 2(S+2)$, where $S=W$ (the width of the $J_{2}$ region) and the factor $2$ comes from the particle and hole orbitals.

\subsection{Transmission probability calculation}

From the solution of $(E-H')\boldsymbol{\psi}=\mathbf{Q}$, the transmitted wave amplitude $\boldsymbol{\Phi}_{2}$ can be obtained, which is a direct product of spin and orbital wavefunctions, 
\begin{align}
\Phi_{2}=
\begin{pmatrix}
a_{+}\\
a_{-}
\end{pmatrix}\otimes \Psi_{2}
\end{align}
We expand the right-moving transmitted wave ($\Phi_{2}$) in the eigenbasis of the right leads,
\begin{align}
\boldsymbol{\Phi}^{(m,\sigma')}_{2}= \sum_{n=1}^{2}\sum_{\sigma=\uparrow,\downarrow}\mathbf{u}^{s}_{\sigma}\otimes\mathbf{u}^{f}_{R,n}(+)\ t_{n\sigma,m\sigma'}
\end{align}
where $\mathbf{u}_{\sigma}^{s}$ and $\mathbf{u}^{f}_{R,n}$ correspond to spin and (right-moving) Bogoliubov fermion wavefunctions, respectively. The coefficients are the elements of the transmission matrix, which are obtained as follows,
\begin{align}
 t_{n\sigma, m\sigma'} = \big[ \mathbf{u}^{s}_{\sigma}\otimes\mathbf{u}^{f}_{R,n}(+) \big]^{\dag} \boldsymbol{\Phi}^{(m,\sigma')}_{2}
\end{align}
After calculating the (4$\times$4) transmission matrix ($\mathbf{t }$), we can obtain the transmission probability ($T$) by taking the trace,
\begin{align}
T(E) = \frac{1}{4}\text{Tr}(\mathbf{t}\mathbf{t}^{\dag})
\end{align}
The $1/4$ factor is for normalization (two factors of $1/2$ account for averaging over the spin and particle-hole spaces, respectively). 
\section{Continuum description of $\tilde{H}^{(m)}$}\label{continuum-model-m-anyon}
In this section, we derive the effective low-energy theory for the lattice model (Eq.~\eqref{Hm-ladder-fermion-version}) that describes the motion of $m$ anyons across the XY junction. To derive that, we divide the Hamiltonian into four parts, $\tilde{H}^{(m)}=H_{f}^{(0)}+H_{\text{dom}}+ V_{f}+ V_{fs}$, where $H_{f}^{(0)}$ is the uniform Kitaev chain Hamiltonian,
\begin{align}
H_{f}^{(0)} = -h_{z}\sum_{i}\big(f_i^{\dag} f_{i+1}^{\dag}+f_i^{\dag}f_{i+1}+ h.c.\big) -J\sum_{i}(1-2n_i) \label{eq-Hf0}
\end{align} 
and $H_{\text{dom}}$ governs the spin dynamics of the domino spin,
\begin{align}
 H_{\text{dom}} = -\sqrt{2}J\tau^{z}_{\text{dom}}-h_z \tau^{x}_{\text{dom}}   
\end{align}
There is a perturbation in the hopping and pairing, acting on the bonds adjacent to the junction (or the site $i=0$), and also an onsite potential at $i=0$,
\begin{align}
V_{f}=\frac{h_z}{2}\bigg[(1+\textrm{i})f_{-1}^{\dagger}f_{0}^{\dagger}+(1-\textrm{i})f_{-1}^{\dagger}f_{0}+(1-\textrm{i})f_{0}^{\dagger}f_{1}^{\dagger} \nonumber\\
+(1-\textrm{i})f_{0}^{\dagger}f_{1}+\textrm{H.c.}\bigg]
+J(1-2f^{\dag}_{0}f_{0})
\end{align}
The domino spin and the fermions are coupled by, 
\begin{align}
V_{fs}= -\frac{h_z}{2}&\bigg[(1+\textrm{i})f_{-1}^{\dagger}f_{0}^{\dagger}+(1-\textrm{i})f_{-1}^{\dagger}f_{0}+(1-\textrm{i})f_{0}^{\dagger}f_{1}^{\dagger}\nonumber\\
&+(1-\textrm{i})f_{0}^{\dagger}f_{1}+\textrm{H.c.}\bigg]\tau^{x}_{\text{dom}}
\end{align}
The first step in taking the continuum approximation involves finding the low-energy modes of the unperturbed system (Eq.~\eqref{eq-Hf0}). In this case, these are the $k\sim 0$ Bogoliubov excitations above $\Delta =2(J-h_{z})$. Near $k=0$, and when $h_{z}<J$, $u_{k\sim 0} = 0$ and $v_{k\sim 0} =1$ (upto a $\mathbb{Z}_{2}$ phase). Therefore, from the relation $f_{k}=u_{k}\gamma_{k}-iv_{k}\gamma^{\dag}_{-k}$, one obtains $\frac{1}{\sqrt{N}}\sum_{k\sim 0}f_{k}e^{-ikx_{i}} = -\frac{i}{\sqrt{N}}\sum_{k\sim 0} \gamma_{k}^{\dag}e^{ikx_{i}}$, This implies $\tilde{f}_{i} = -i\tilde{\gamma}_{i}^{\dag}$, where we put tilde on the operators to distinguish their $k\sim 0$ (soft) mode from the lattice fermion operators. The latter also involve rapidly oscillating parts, $f_{i}=\tilde{f}_{i} + \frac{1}{\sqrt{N}}\sum_{|k| > \Lambda_{0}} f_{k}e^{-ikx_{i}}$, where $\Lambda_{0}$ is the cut-off beyond which the energy dispersion of $H_{f}^{(0)}$ is no longer quadratic. In the long-wavelength limit, we can safely ignore such large-momentum contributions (beyond $\Lambda_{0}$). Thus, the low-energy Bogoliubov modes are {\it hole-like} excitations in our regime of interest.

Next, we normalize the lattice operators to get the continuum field operators,
\begin{align}
\psi(x_{i}) = \frac{1}{\sqrt{a_{0}}}\tilde{\gamma}_{i}
\end{align}
The non-interacting Kitaev chain Hamiltonian (Eq.~\eqref{eq-Hf0}) can be approximated as follows, 
\begin{align}
H_{f}^{(0)} = -\int dx\ \psi^{\dag}(x)\bigg(\Delta_{m} +\frac{\partial_{x}^{2}}{2M_{m}}\bigg)\psi(x)
\end{align}
where $\tilde{J}=Ja_{0}$, $\tilde{h}_{z}=h_{z}a_{0}$ are the effective coupling strengths that survive in the continuum limit ($J,\ h_{z}\rightarrow \infty$, $a_{0}\rightarrow 0$). The mass gap and the band mass of the quasiparticle are given by $\Delta_{m} = 2Ja_{0}\big(1-h/J\big)$ and $M_m=\Delta_{m}/4\tilde{h}_{z}\tilde{J}$.\\
\indent Next, we find what happens to $V_{f}$ in the continuum limit: 
\begin{align}
& h_{z}\big[(1-i)f_{0}^{\dag}f_{1}^{\dag}-(1+i)f_{0}^{\dag}f_{-1}^{\dag} \big]\nonumber\\
&\approx h_{z}\big[-(1-i)\gamma_{0}\gamma_{1}+(1+i)\gamma_{0}\gamma_{-1}\big] \ \ \ 
\nonumber\\
&=-\tilde{h}_{z}\big[(1-i)\psi(0)\psi(a_{0})-(1+i)\psi(0)\psi(-a_{0})\big]\nonumber\\
&\approx -2a_{0}\tilde{h}_{z}\psi(x)\partial_{x}\psi(x)\big|_{x=0} + \mathcal{O}(a_{0}^{2})
\end{align}
Since there is no sum over lattice sites, this term vanishes in the continuum limit, $a_{0}\rightarrow 0$. For the hopping part, a finite contribution remains,
\begin{align}
&h_{z}\big[(1-i)f_{-1}^{\dag}f_{0}+(1-i)f_{0}^{\dag}f_{1}+h.c.\big]\nonumber\\
&\approx\tilde{h}_{z}[(1-i)\psi(-a_{0})\psi^{\dag}(0)+(1-i)\psi(0)\psi^{\dag}(a_{0})+\textrm{H.c.}]\nonumber\\
&\nonumber \approx -4\tilde{h}_{z}\psi^{\dag}(0)\psi(0) + 2ia_{0}\tilde{h}_{z}\bigg[\psi^{\dag}\partial_{x}\psi-(\partial_{x}\psi^{\dag})\psi\bigg]_{x=0}+\\
&\mathcal{O}(a_{0}^{2})
\end{align}
Only the first term survives in the limit $a_{0}\rightarrow 0$ and therefore we obtain,
\begin{align}
 V_{f}\approx \Delta_{m} \psi^{\dag}(x=0)\psi(x=0)    
\end{align}
and $V_{fs}$ simplifies to the following expression,
\begin{align}
V_{fs} = \eta h_{z}\psi^{\dag}(0)\psi(0)\tau^{x}_{\text{dom}}
\end{align}
where $\eta$ is a non-universal cut-off ($\Lambda_{0}$) dependent factor. Combining all these pieces, we obtain, 
\begin{align}
H^{(m)} = &-\int dx\ \psi^{\dag}(x)\bigg[\Delta_{m}(x) +\frac{\partial_{x}^{2}}{2M_{m}}\bigg]\psi(x)  -\sqrt{2}J\tau^{z}_{\text{dom}}\nonumber\\
&-h_z \tau^{x}_{\text{dom}}+ \eta h_{z} \psi^{\dag}(0)\psi(0)\tau^{x}_{\text{dom}} \label{continuum-model-bicolor}
\end{align}
where $\Delta_{m}(x)=\Delta_{m}[1-\delta(x)]$ is the inhomogeneous mass gap of the Bogoliubov soft modes ($\psi$). Note that these excitations face a Dirac delta barrier at the origin ($x=0$) and the barrier strength depends on the combination $\Delta_{m}+\eta h_{z}\langle \tau^{x}_{\text{dom}}\rangle$, which depends on the ratio $h_{z}/J$. At the level of a mean-field approximation,  $\langle \tau^{x}_{\text{dom}}\rangle =\tilde{h}_{z}/\sqrt{\tilde{h}_{z}^{2}+2J^{2}}$, where $\tilde{h}_{z}=h_{z}(1-2\langle \gamma_{0}^{\dag}\gamma_{0}\rangle)$. Now, $\langle \gamma^{\dag}_{0}\gamma_{0}\rangle=1-\langle f^{\dag}_{0}f_{0}\rangle$. When $h_{z}\sim 0$, there is no onsite energy term for $n^{f}_{i=0}$ and the hopping that connects $i=0$ to the rest of the chain is only of $O(h_{z})$, therefore negligible. So, we expect $\langle f^{\dag}_{0}f_{0}\rangle_{h\rightarrow 0} \approx 1/2$. For $i\neq 0$, $\langle f^{\dag}_{i}f_{i}\rangle_{h\rightarrow 0}\approx 0 $, because the onsite potential is positive (see Eq.~\eqref{eq-Hf0}). Naturally, when $h_{z}$ becomes nonzero (but still small), $\langle f^{\dag}_{0}f_{0}\rangle < 1/2$, due to hopping between site $i=0$ and the rest of the system. As a result, $\langle \tau^{x}_{\text{dom}}\rangle$ switches to a negative number. This is also confirmed in our $m$ anyon wavepacket dynamics calculation (see Fig.~(3) of the main text) of the lattice model Eq.~\eqref{Hm-ladder-fermion-version}. As shown in Fig.~\ref{fig:domino-spin-avg}, the domino spin remains rigidly aligned along the $z$ direction, but once the $m$ anyon hits the junction (around $Jt\approx 400$), $\langle \tau^{x}_{\text{dom}}\rangle$ becomes negative. The height of the delta barrier therefore decreases due to the combined effect of (i) reduction in onsite energy ($\Delta_{m}$) at $i=0$ and (ii) because $\langle \tau^{x}_{\text{dom}}\rangle$ becomes negative as $h_{z}$ increases.

\begin{figure}
    \centering
    \includegraphics[width=0.5\linewidth]{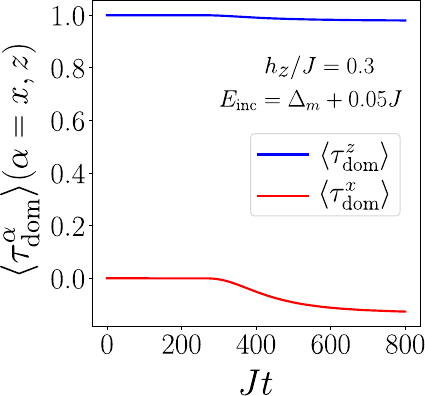}
    \caption{Time evolution of the expectation values of domino spin components ($\langle \tau^{\alpha}_{\text{dom}}\rangle(t)$, for $\alpha=x,z$). The numerical data is for incident $m$ anyon energy, $E_{\text{inc}}=\Delta_{m}+0.05J$, and magnetic field strength, $h_{z}/J=0.3$. The $m$ anyon wave packet evolution is shown in the main text (see Fig.(3)) for the same data points.}
    \label{fig:domino-spin-avg}
\end{figure}

\section{Mapping $e$ anyon dynamics to a 1d $Z_{2}$ gauge theory} \label{Z2-gauge-mapping-e-dynamics}
While the $e$ particle can move freely within the $\mathcal{N}$ region, its behavior becomes complicated upon entering the $\mathcal{S}$ region. Initially, right next to the junction, the hopping of the $e$ charge induces quantum fluctuations in the dimer ($H_{d}$) and domino ($\tau^{x}_{\text{dom}}$) sectors (see Eq.~\eqref{eff-action-sigmax-eq}). Quantum fluctuations in the domino sector are suppressed if $h_{x}<J$. In the XYJC model, dimer spins can flip freely since these excitations are energetically degenerate in the zero-field limit, resulting in an almost negligible energy cost for flipping a dimer. Hence, the $e$ charge can hop easily by flipping a dimer pair and cross the $\mathcal{E}^{+}$ region (see Fig.~(3) of the main text). Across the $\mathcal{S}^{+}$ region, the $e$ charge hopping generates pair fluctuations in the $m$ sector. We will focus solely on the effects of $m$ charge pairs on the dynamics of an $e$ particle, as this process occurs throughout the $\mathcal{S}^{+}$ region, leading to a significant influence on the transport properties of the $e$ particle.

\indent Consider the part of the effective spin Hamiltonian that describes $e$ charge dynamics in the $\mathcal{S}^{+}$ region,
\begin{align}
H_{\mathcal{S}^{+}}^{(e)} &= -J\sum_{v\in r_{e}} \tau^{z}_{v} -J\sum_{v\in r_{e},\ \sigma=\pm} \tau^{z}_{v+1/2, \sigma} \nonumber\\
&\ \ \ \ \ \ \ \ -h_{x}\sum_{v\in r_{e}}\tau^{x}_{v}(\tau^{x}_{v+1/2,+}\tau^{x}_{v+1/2,-})\tau^{x}_{v+1}
\end{align}
Here, $v$ denotes the location of the $A_{v}^{z}$ operators (or the static $e$ anyon excitations) and $(v+1/2, \pm)$ denote the two plaquette centers that share a common link $(v,v+1)\in r_{e}$. This Hamiltonian describes the motion of gapped $e$ anyon excitations in the vacuum of $m$ charges, where their movement induces pair fluctuations of the $m$ particles. We disregard the $h_{z}$ coupling, which induces fluctuations solely in the $m$ sector, and focus on examining the action of the $h_{x}$ Zeeman field on the otherwise stationary $e$ and $m$ anyons. Conclusions obtained from this approximation will also remain valid when $h_{z}$ is finite but $h_{z}< J$ (so that the $m$ anyon gap $\sim (J-h_{z})$ remains finite). 

The $h_{x}$ field creates $m$ charges in pairs, at the plaquette centers labeled by $(v+1/2, \pm)$. Thus, the $h_{z}=0$ low-energy subspace for a given bond along the ladder rung is two-dimensional and comprises the states,
\begin{align}
\ket{\Uparrow} = \ket{\tau^{z}_{v+\frac{1}{2},+}=+1} \otimes \ket{\tau^{z}_{v+\frac{1}{2},-}=+1} \nonumber\\
\ket{\Downarrow} = \ket{\tau^{z}_{v+\frac{1}{2},+}=-1} \otimes \ket{\tau^{z}_{v+\frac{1}{2},-}=-1}
\end{align}
Consider the following operators defined on the bonds $(v,v+1)\in r_{e}$,
\begin{align}
X_{v, v+1} = \frac{1}{2}\Big(\tau^{z}_{v+\frac{1}{2},+} + \tau^{z}_{v+\frac{1}{2},-}\Big)\ ,\ Z_{v, v+1} = \tau^{x}_{v+\frac{1}{2},+}\tau^{x}_{v+\frac{1}{2},-} 
\end{align}
These operators behave as Pauli spins within the low-energy subspace of $\ket{\Uparrow}$ and $\ket{\Downarrow}$, i.e., $(X_{vv'})^{2}=1=(Z_{vv'})^{2}$ and $[Z_{vv'}, X_{vv'}]_{+}=0$ (where $v'=v+1$). In terms of these {\it bond-spins}, our Hamiltonian becomes a 1d $\mathbb{Z}_{2}$ gauge theory,
\begin{align}
H_{\mathcal{S}^{+}}^{(e)} = -J\sum_{v} \tau^{z}_{v} - 2J\sum_{v} X_{v,v+1} -h_{x}\sum_{v} \tau^{x}_{v} Z_{v,v+1}\tau^{x}_{v+1} 
\end{align}
The operator, $G_{v} = X_{v-1,v}\tau^{z}_{v}X_{v,v+1}$ is a symmetry of the above Hamiltonian. The bond-spins (and site-spins) correspond to the $\mathbb{Z}_{2}$ gauge fields (and $\mathbb{Z}_{2}$ matter fields) of the gauge theory. Restricting ourselves to a specific symmetry sector is equivalent to imposing the Gauss law constraint. The zero field sector corresponds to $G_{v}=+1\ \forall \: v$ (finite $h_{x}$ cannot change the symmetry sector). 

One-dimensional gauge theories with matter show confinement. The gauge field string connecting two matter ($e$) charges will cost an energy that scales as the length of the string (or the separation between two $e$ charges). The same is also true for a single $e$ charge motion. The string tension thus grows as $\sim 4Jl$ ($l$= length of the string).

Thus, the $e$ particle's motion in the $\mathcal{S}^{+}$ region is similar to that of a quantum particle under the influence of a linearly increasing potential, $V(x) =0,\ \text{for }x<0$, and $V(x) = \lambda x, \ \ \text{for }x\geq 0$. This leads to negligible transmission and significant reflection back into the $\mathcal{N}$ region, when $h_{x}<J$. This explains the large opacity of the XYJC to the $e$ charge.  

\section{Effective action of $h_z$ Zeeman field (for small $\delta \theta = \pi/2-\theta$)} \label{theta-neq-pi2-h-action}

First, we write down the expression of the general energy eigenstate of the $h_{z}=0$ model, restricted to the $e$ charge-free sector ($A_{v}^{z}=+1$),
\begin{widetext}
\begin{align}
&\big| \lbrace r^{h}_{d}\rbrace'_{d\in \mathcal{E}}, \lbrace r_{d}\rbrace_{d\in \mathcal{E}},\lbrace r_{p}\rbrace_{d\in (\mathcal{N}^{+},\mathcal{S}^{+})}\big\rangle= \bigg[1+ (-1)^{\sum_{p\in (\mathcal{N}^{+}, \mathcal{S}^{+})}r_{p}}\prod_{d\in \mathcal{E}}H_{d}\bigg]\nonumber\\
&\ \ \times\prod_{d\in \mathcal{D}}\bigg[1+\frac{(-1)^{r_{d}}}{\sqrt{2}}(1+H_{d}\cos{\theta})^{-\frac{1}{2}}D_{d}\bigg]\big(1+(-1)^{r^{h}_{d}}H_{d}\big)\prod_{p\in (\mathcal{N}^{+}, \mathcal{S}^{+})}(1+(-1)^{r_{p}}B^{x}_{p})\ket{\uparrow\uparrow....\uparrow}_{z}
\end{align}
\end{widetext}
Since our focus is on scattering phenomena, which inherently involve open boundary conditions, the global even dimer-parity projection will be omitted in the subsequent calculations.

(A) \textit{Action of $\sigma^{z}_{l}$ far into the $\mathcal{N}^{+}$, $\mathcal{S}^{+}$ regions}:  In this case, $\sigma^{z}_{l}$ creates two $m$ charges in the neighboring plaquettes in both the north and the south regions. Hence, as before, the effective action can be expressed as $\sigma^{z}_{l} \rightarrow \tau^{x}_{p}\tau^{x}_{p'}$, where the dual link $\langle pp'\rangle$ cuts the link $l$ of the direct lattice.

(B) {\it Action of $\sigma^{z}$ on the boundary between ($\mathcal{N}^{+}$, $\mathcal{S}^{+}$) and $\mathcal{E}$ regions}:  As before, $\sigma^{z}_{l^{-}}$, which acts on the boundary between the $\mathcal{N}^{+}$ and $\mathcal{E}$ regions, excites a $m$ anyon (flips a $B_{p}^{x}$ on the left), a dimer-flip, and converts $D_{d}(\theta)$ to $D'_{d}(\theta)$ (i.e., creates a linear superposition of state having both the domino flipped and unflipped),
\begin{widetext}
\begin{align}
&\sigma^{z}_{l^{-}}\big| \lbrace r^{h}_{d}\rbrace_{d\in \mathcal{E}}, \lbrace r_{d}\rbrace_{d\in \mathcal{E}},\lbrace r_{p}\rbrace_{d\in (\mathcal{N}^{+},\mathcal{S}^{+})}\big\rangle =\bigg(1+(-1)^{r_{d}^{h}+1}H_{d}\bigg) \prod'_{d'\in \mathcal{E}}\bigg(1+(-1)^{r_{d'}^{h}}H_{d'}\bigg)\nonumber\\
&\ \ \ \ \ \ \ \ \ \ \ \ \ \times \bigg[1+\frac{(-1)^{r_{d}+1}}{\sqrt{2}}(1-H_{d}\cos{\theta})^{-\frac{1}{2}}D'_{d}(\theta)\bigg] \prod'_{d'\in \mathcal{E}}\bigg[1+\frac{(-1)^{r_{d'}}}{\sqrt{2}}(1+H_{d}\cos{\theta})^{-\frac{1}{2}}D_{d'}(\theta)\bigg]\nonumber\\
&\ \ \ \ \ \ \ \ \ \ \ \times \bigg(1+(-1)^{r_{p_{l}}+1}B^{x}_{p_{l}}\bigg) \prod'_{p'\in \mathcal{N}^{+},\mathcal{S}^{+}}\bigg(1+(-1)^{r_{p'}}B^{x}_{p'}\bigg) \big | \uparrow\uparrow....\uparrow\big \rangle_{z} \label{sz-action1}
\end{align}    
\end{widetext}
Here, we used the identity $\sigma^{z}_{l}(1+H_{d}\cos{\theta})^{-\frac{1}{2}}=(1-H_{d}\cos{\theta})^{-\frac{1}{2}}\sigma^{z}_{l}$. This can be proved if we assume $\theta\lesssim \pi/2$, and do a series expansion of the square root function. Next, similar to the procedure outlined for the XYJC in Sec.~\ref{field-induced-dynamics-m-anyon-supp}, we calculate the following overlaps, 
\begin{align}
&\bra{r_{d}^{h}+1, r_{p_{l}}+1, r_{d}}\ \sigma^{z}_{l^{-}}\ \ket{r_{d}^{h}, r_{p_{l}}, r_{d}}\nonumber\\
&\ \ =\frac{1}{2}\bigg(1-\frac{(-1)^{r^{h}_{d}+1}}{2\sin{\theta}}\big\langle H^{x}_{d}D_{d}(\theta)D_{d}'(\theta)\big\rangle\bigg) \label{overlap1-1}
\end{align}
\begin{align}
&\bra{r_{d}^{h}+1, r_{p_{l}}+1, r_{d}+1}\ \sigma^{z}_{l^{-}}\ \ket{r_{d}^{h}, r_{p_{l}}, r_{d}}\nonumber\\
&\ \ = \frac{1}{2}\bigg(1+\frac{(-1)^{r^{h}_{d}+1}}{2\sin{\theta}}\big\langle H^{x}_{d}D_{d}(\theta)D_{d}'(\theta)\big\rangle\bigg) \label{overlap2-1}
\end{align}
To obtain the R.H.S., we used $(1+H^{x}_{d}\cos{\theta})^{-\frac{1}{2}}(1-H^{x}_{d}\cos{\theta})^{-\frac{1}{2}} \approx 1/\sin{\theta}$ (which also depends on the restriction that $\theta$ is close to $\pi/2 $). Now, the expectation value $\big\langle H^{x}_{d}D_{d}(\theta)D_{d}'(\theta)\big\rangle$ for the $z$-polarized state is $-2i\sin{\theta}$. Note that both limits $\theta=0$ and $\pi/2$ match with expectations. Then Eqs.~\eqref{overlap1-1}, \eqref{overlap2-1} can be further simplified, and we finally obtain,
\begin{align}
\bra{r_{p_{l}}+1, r_{d}^{h}+1, r_{d}}\ \sigma^{z}_{l^{-}}\ &\ket{r_{d}^{h}, r_{p_{l}}, r_{d}}\nonumber\\
&= \frac{1}{\sqrt{2}}\exp{\big[-i\frac{\pi}{4}(-1)^{r^{h}_{d}}\big]}\\
\bra{r_{p_{l}}+1, r_{d}^{h}+1, r_{d}+1}\ \sigma^{z}_{l^{-}}\ &\ket{r_{d}^{h}, r_{p_{l}}, r_{d}}\nonumber\\
&= \frac{1}{\sqrt{2}}\exp{\big[i\frac{\pi}{4}(-1)^{r^{h}_{d}}\big]}
\end{align}
These amplitudes are independent of the parameter $\theta$. Similar calculations can be done for $\sigma^{z}_{l^{+}}$ (the link at the boundary of $\mathcal{E}$ and $\mathcal{S}^{+}$) and the final result is that the effective action of $\sigma^{z}_{l^{-}}$ and $\sigma^{z}_{l^{+}}$ in terms of the dual spins have the same form as Eqs.~\eqref{eff-sigmazl}, \eqref{eff-sigmazr}. At the equator, $\sigma^{z}_{\text{eq}}$ only flips the domino spin, which is the same as XYJC, so Eq.~\eqref{eff-sigmaz-eq} also holds here.

Collecting all these terms, we finally obtain,
\begin{align}
\tilde{H}_{m}^{(\theta=\pi/2^-)} &= - \Delta_{d}(\theta) \tau^{z}_{\text{dom}} -h_{z} \tau^{x}_{\text{dom}}  -\frac{1}{2}\Delta_{v}(\theta)\tau^{z}_{0}\tau^{z}_{\text{dom}}\nonumber\\
&\ \ \ \ - h_{z}\sum_{i\not \in \lbrace -1, 0\rbrace} \tau^{x}_{i}\tau^{x}_{i+1} -J\sum_{i \neq 0} \tau^{z}_{i} \nonumber\\
&\ \ \ \ -\frac{h_{z}}{2} \bigg[ \big(\tau_{-1}^x \tau_{0}^x + \tau_{0}^x \tau_{+1}^x\big) (\tau^x_{\textrm{dom}}+\mathds{1}) \nonumber\\
&\ \ \ \ \ \ \ \ \ \ +\big(\tau_{-1}^x \tau^y_{0}-\tau^y_{0}\tau_{+1}^x\big)(\tau_{\textrm{dom}}^x-\mathds{1}) \bigg].\label{m-anyon-eff-H-theta}
\end{align}

\bibliography{refs}

\begin{thebibliography}{68}
\bibitem [{jun()}]{junction_notation}%
  \BibitemOpen
  \href@noop {} {}\bibinfo {note} {The $\mathcal{N}, \mathcal{S}$ nomenclature
  for the two regions is inspired by considering the toric code on a manifold
  of genus 0 (a sphere such as the Earth). The junction line is thus called the
  equator $\mathcal{E}_q$.}\BibitemShut {Stop}%
\bibitem [{not()}]{note-pbc}%
  \BibitemOpen
  \href@noop {} {}\bibinfo {note} {For a periodic lattice, there are two
  junctions, each of length $L_{y}$, where the twisted and untwisted regions
  meet.}\BibitemShut {Stop}%
\end{thebibliography}

\end{document}